\documentclass[AMA,STIX2COL]{WileyNJD-v2} 

\usepackage{moreverb}
\usepackage{subfigure}
\usepackage{mhchem}
\usepackage{lipsum}

\newcommand{\LB}{\linebreak}

\newif\ifcomments
\commentstrue

\newcommand\BibTeX{{\rmfamily B\kern-.05em \textsc{i\kern-.025em b}\kern-.08em
T\kern-.1667em\lower.7ex\hbox{E}\kern-.125emX}}

\articletype{Feature Article}%

\received{<day> <Month>, <year>}
\revised{<day> <Month>, <year>}
\accepted{<day> <Month>, <year>}

\begin{document}


\title{PECVD and PEALD on polymer substrates (Part II): Understanding and tuning of barrier and membrane properties of thin films}

\author[1]{Teresa de los Arcos}
\author[2]{Peter Awakowicz}
\author[3]{Marc B\"oke}
\author[4,5]{Nils Boysen}
\author[6]{Ralf Peter Brinkmann}
\author[7]{Rainer Dahlmann}
\author[4,5]{Anjana Devi}
\author[6]{Denis Eremin}
\author[7]{Jonas Franke}
\author[8]{Tobias Gergs}
\author[2]{Jonathan Jenderny}
\author[6]{Efe Kemaneci}
\author[9]{Thomas D. K\"uhne}
\author[7]{Simon Kusmierz}
\author[8]{Thomas Mussenbrock}
\author[10]{Jens Rubner}
\author[11]{Jan Trieschmann}
\author[10,12]{Matthias Wessling}
\author[1]{Xiaofan Xie}
\author[4]{David Zanders}
\author[9]{Frederik Zysk}
\author[1]{Guido Grundmeier*}

\address[1]{\orgdiv{Technical and Macromolecular Chemistry}, \orgname{Paderborn University}, \orgaddress{Warburger Str. 100, 33098 Paderborn, \country{Germany}}}

\address[2]{\orgdiv{Chair of Electrical Engineering and Plasma Technology}, \orgname{Ruhr University Bochum}, \orgaddress{Universitaetsstraße 150, 44801 Bochum, \country{Germany}}}

\address[3]{\orgdiv{Chair of Experimental Physics II: Physics of Reactive Plasmas}, \orgname{Ruhr University Bochum}, \orgaddress{Universitaetsstraße 150, 44801 Bochum, \country{Germany}}}

\address[4]{\orgdiv{Inorganic Materials Chemistry}, \orgname{Ruhr University Bochum}, \orgaddress{Universitaetsstraße 150, 44801 Bochum, \country{Germany}}}

\address[5]{\orgname{Fraunhofer IMS}, \orgaddress{47057 Duisburg, Germany}}

\address[6]{\orgdiv{Institute of Theoretical Electrical Engineering}, \orgname{Ruhr University Bochum}, \orgaddress{Universitaetsstraße 150, 44801 Bochum, \country{Germany}}}

\address[7]{\orgdiv{Institute for Plastics Processing}, \orgname{RWTH Aachen University}, \orgaddress{Seffenter Weg 201, 52074 Aachen, \country{Germany}}}

\address[8]{\orgdiv{Chair of Electrodynamics and Plasma Technology}, \orgname{Ruhr University Bochum}, \orgaddress{Universitaetsstraße 150, 44801 Bochum, \country{Germany}}}

\address[9]{\orgdiv{Chair of Theoretical Chemistry}, \orgname{Paderborn University}, \orgaddress{Warburger Str. 100, 33098 Paderborn, \country{Germany}}}

\address[10]{\orgname{DWI -- Leibniz-Institute for Interactive Materials}, \orgaddress{Forckenbeckstr. 50, 52074 Aachen, \country{Germany}}}

\address[11]{\orgdiv{Theoretical Electrical Engineering}, \orgname{Kiel University}, \orgaddress{Kaiserstraße 2, 24143 Kiel, \country{Germany}}}

\address[12]{\orgdiv{Chemical Process Engineering AVT.CVT}, \orgname{RWTH Aachen University}, \orgaddress{Forckenbeckstr. 51, 52074 Aachen, \country{Germany}}}

\authormark{G. Grundmeier \textsc{et al}}
\corres{*G. Grundmeier, Technical and Macromolecular Chemistry, Paderborn University, 33098 Paderborn, Germany\\ \email{guido.grundmeier@uni-paderborn.de}}

\abstract[Abstract]{This feature article presents insights concerning the correlation of PECVD and PEALD thin film structures with their barrier or membrane properties. While in principle similar precursor gases and processes can be applied, the adjustment of deposition parameters for different polymer substrates can lead to either an effective diffusion barrier or selective permeabilities. In both cases the understanding of the film growth and the analysis of the pore size distribution and the pore surface chemistry is of utmost importance for the understanding of the related transport properties of small molecules. In this regard the article presents both concepts of thin film engineering and analytical as well as theoretical approaches leading to a comprehensive description of the state of the art in this field. Moreover, based on the presented correlation of film structure and molecular transport properties perspectives of future relevant research in this area is presented.}


\maketitle

\section{Introduction (general state of the art)}\label{sec1}

\ifcomments
\else\fi

The broad range of coating properties resulting from the high variability of plasma-enhanced chemical vapor deposition (PECVD) and atomic layer deposition (PEALD) processes has opened up a broad spectrum of applications for these processes. In particular, the deposition of layered systems on plastics gives plastic products new functionalities, with the help of which a wealth of products can meet new requirements.

The barrier properties of coatings, which can be adjusted via the process parameters, offer important perspectives for product groups such as plastic food packaging. Plasma-assisted layer systems are certified to have a high recyclability and can therefore provide solutions to the current worldwide demands for a circular economy for plastics. Additionally, layer systems produced by PECVD and PEALD can be selectively adjusted. The versatibility of PECVD and PEALD opens the possibility to create sophisticated layer architectures leading to selective barriers which, deposited on highly permeable membranes, allow targeted separation of mixed gases.

This paper represents part of the research developed within the frame of the transregional project SFB-TR 87 ("Pulsed high power plasmas for the synthesis of nanostructured functional layers"), which aims to understand mass transfer processes through coated plastics and membranes. To this end, experimental methods are contrasted with simulative methods and discussed.

\subsection{Barrier films}\label{subsec11}

In the last decades, plastics have become essential in numerous applications. Their affordability and the ease of adjusting their properties across a wide range were the decisive factors. However, their relatively high permeability to gases and vapors posed a challenge for applications like food packaging and electronic protection \cite{GiacintiBaschetti.2020, LeFloch.2018}. Various methods have been developed to enhance barrier properties, such as employing multilayer films with barrier polymers like ethylene vinyl alcohol (EVOH) or blends with polyamide (PA), or utilizing oxygen absorbers \cite{McKeen.2017, Gaikwad.2018, Bauer.2021}. Unfortunately, most of these techniques require a significant amount of additives, compromising the recyclability of the base polymer. In contrast, thin-film coatings, such as aluminum-oxide (AlO$_x$) or silicon-oxide (SiO$_x$), overcome this drawback as they only constitute a fraction of the total mass. Hence, they do not hinder the recycling process while effectively providing excellent barrier properties to their substrates \cite{Bauer.2021, Wilski_2020}.

The focus on recyclability has sparked a growing interest in thin film barrier coatings, although their development has mostly relied on empirical or experiential approaches. Therefore, significant advances in process control for the generation of PECVD barrier coatings are necessary to facilitate the up-scaling and wide-spread use of this technology. For this, in-depth understanding of the relationship between process parameters and the barrier coating’s properties must be gathered. This includes the analysis of porosity of the coating, which exerts considerable influence on the resulting permeability as well as the consideration of the relationship between process parameters and resulting coating properties. This work provides an overview of investigations to cover both of these aspects.

\subsection{Membrane films}\label{subsec12}


Membranes are thin layers of polymers, ceramics, or metals possessing semi-permeable properties. These membranes play a crucial role in various industrial processes as a fundamental operation in process engineering. Their remarkable energy efficiency sets membrane separation processes apart from traditional thermally driven separation processes. In 2016, Sholl and Lively highlighted the significant benefits of non-thermal purification methods, such as membrane-based processes, which substantially reduce energy consumption, emissions, and pollution \cite{sholl2016seven}.

The driving force of membrane processes is a difference in chemical potential. In the subsequent applications, this potential is mainly applied by a pressure difference between upstream (feed) and downstream (permeate) compartment. Dependent on the size of the pores or free volumina present in the membrane, they are categorized into microfiltration, ultrafiltration, nanofiltration, gas separation and reverse osmosis membranes by decreasing pore size. Currently, membranes find applications in water treatment, food and beverage processing, gas separation, and medical fields. Nevertheless, their potential extends beyond these specific areas, as they hold the key to addressing numerous global challenges of the 21st century. However, improvements in production, process costs, and intrinsic membrane properties are necessary to increase their market share in industrial applications.

In gas separation, extensive research has been conducted over the past few decades to enhance permeability and selectivity by developing novel materials through polymer synthesis. This endeavor has led researchers to explore emerging membrane material classes like mixed matrix membranes (MMM), polymers with intrinsic microporosity (PIM), carbon molecular sieves (CMS), and others. These materials have garnered significant attention in scientific journals, pushing the boundaries of membrane properties. However, achieving these exceptional results has made membranes' synthesis and production processes increasingly complex.

Consequently, scaling up fabrication becomes a challenge, often hindering testing these materials in pilot-scale scenarios. Additionally, non-ideal effects can cause considerable changes in permeance or selectivity of the polymers over time. As a result, although promising membrane materials have been developed over the past two decades, they have not yet entered the market.

In addition to the materials mentioned above, there is a possibility of creating thin selective coatings through scalable plasma processes \cite{roualdes20171}. Among these processes, PECVD has been extensively studied over the years \cite{li1999gas}. It allows customized coatings to be produced with precise control over mechanical and chemical properties while achieving thin coating thicknesses in the nanometer range. One application of PECVD is the formation of gas separation membranes using organosilica layers derived from hexamethyldisiloxane (HMDSO) as a precursor \cite{li1999gas,lo2010control,kafrouni2010synthesis,coustel2014insight,nagasawa2015microporous,kleines2020enhancing,kleines2022evaluation}. Inorganic \cite{kafrouni2009synthesis,tsuru20112,nagasawa2013characterization,haacke2015optimization} or organic substrate membranes \cite{li1999gas,kleines2020enhancing,kleines2021structure,kleines2022evaluation,roualdes1999gas,roualdes2002gas,bosc2003sorption,ngamou2013plasma,charifou2016siox,wang2017gas}support these organosilica layers. PECVD offers the advantage of combining the strengths of different membrane materials resulting in high permeance and selectivity membranes for small gas molecules \cite{nagasawa2013characterization}.

Moreover, PECVD is already being used industrially and is scalable in roll-to-roll processes, making it easily adaptable for large-scale membrane production \cite{izu1995roll}. Since the resulting selective layer's properties depend on the plasma parameters \cite{kleines2021structure,kleines2022evaluation}, it is possible to fabricate membranes for various separation tasks on the same production line by simply adjusting the plasma parameters.

Initially, highly-selective organosilica membranes could only be deposited on expensive ceramic support membranes. By utilizing cost-effective organic membranes as a mechanical support in combination with an organosilica layer, membranes fabricated by PECVD would gain significant attraction for use in industrial processes. This work summarizes the recent development of PECVD fabricated membranes on organic supports. It illustrates the dependence of the plasma parameters on the membrane's gas separation properties and the optimization process leading to outstanding membrane properties for the separation of Helium from gas mixtures.\\

\section{Fundamentals of permeation through films (simulation), pores in thin plasma films}\label{sec2}

\subsection{Classification of transport through pores} \label{subsec21}

\begin{figure}[h]
\centering
\includegraphics[width=0.48\textwidth]{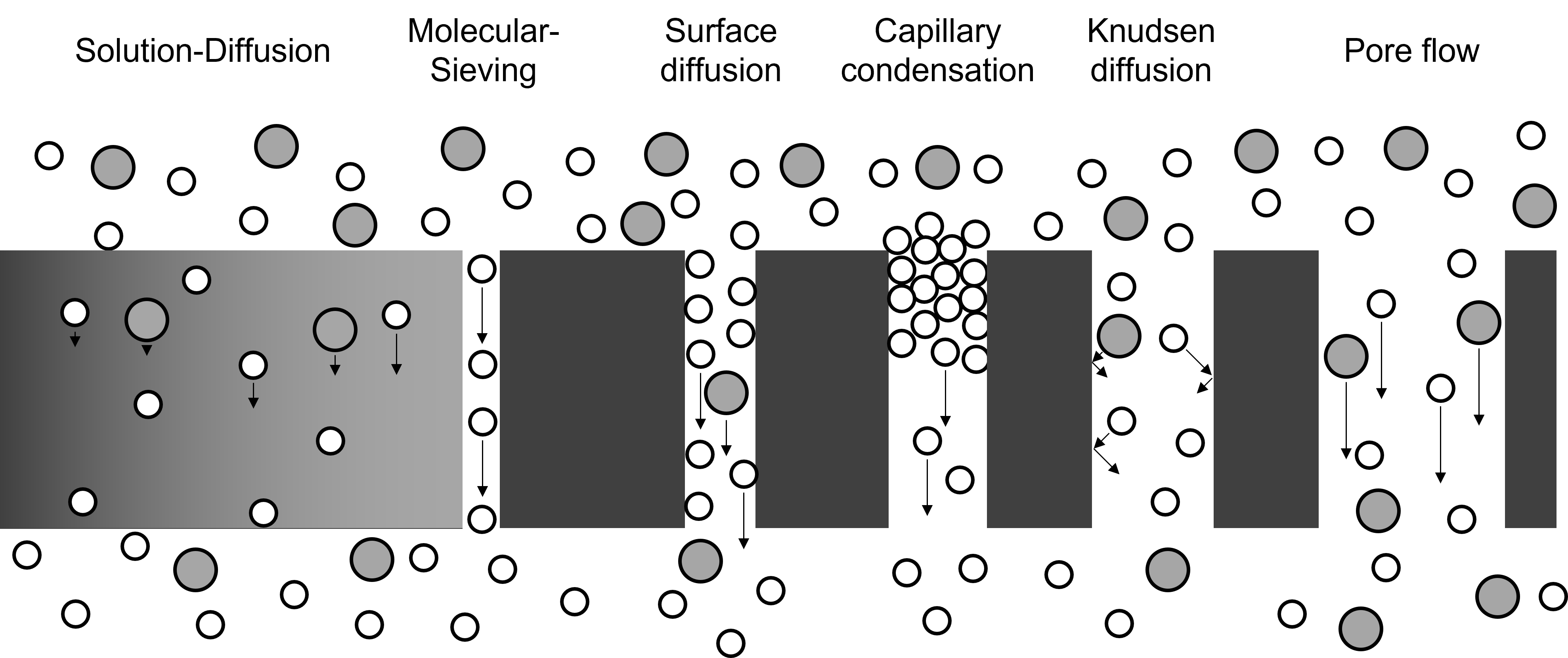}
\caption{Transport phenomena in coatings and membranes through different sized pores.}
\label{fig:DiffMechanisms}
\end{figure}

The control of permeation rates of distinct species through plasma-deposited films is one of the most important properties for the applicability of the deposited films. Figure \ref{fig:DiffMechanisms} displays different transport phenomena in dependence on the pore sizes in the film. Two models describe the permeation mechanisms in plasma-deposited films: the pore-flow model and the solution-diffusion model.

In the pore-flow model (or Poiseuille flow), the number and size of the pores and the interaction of the permeating molecules with the solid interface of the pores control the transport of gas molecules through the films. Pore-flow  dominates the permeation through pores when the pore radius is larger than the mean free path of the gas molecule. In this case, the gas molecules do not interact with the solid interface; therefore, no separation is achieved. In the Knudsen regime, the pore wall interacts with the molecules, hindering their movement.

The diffusion coefficient in the Knudsen regime is dependent on the pore diameter, absolute temperature, and the molecular weight of the molecule. Hence, the selectivity $\alpha_{i,j}$ of molecule i to j of a membrane in the Knudsen regime is only dependent on the molecular weight of the molecules ($m_i$ and $m_j$) and defined as
\begin{equation}
    \alpha_{i,j}=\sqrt{\frac{m_j}{m_i}}.
\end{equation}
In order to have Knudsen flow as the predominant mechanism, the pore radii must be less than 500~\AA since the mean free path of gases at atmospheric pressure is in the range of 500-2000 \AA. There, the molecules have more collisions with the pore walls than with other gases. Each molecule moves independently with an average velocity differing from other species only by the molecular mass.

For pore diameters below 50-100 \AA, surface adsorption and diffusion can occur. In this case, a significant amount of gases adsorb onto the pore walls. The adsorbed molecules are mobile and move to the permeate side by surface diffusion. Depending on the condensability of gases, the adsorbed gases can restrict or block the permeation of the noncondensable gas. This transport phenomenon is called capillary condensation and strongly favors the permeation of the more condensable gas.

Reducing the pore diameter of a microporous membrane to 5-10~\AA, the pores begin to separate the gases by a molecular size sieving effect. Differences in molecular size of only 0.2~\AA~ can change the permeance by orders of magnitude.\cite{baker2012membrane}

In contrast to the pore-flow model, the solution-diffusion model is not based on convective flow through pores but on molecules that dissolve in a nonporous film and then diffuse through it down a concentration gradient. It is mainly used to model the permeation through dense polymer films. The permeability $P_i$ of a gas i through a material is described by the product of its diffusivity $D_i$ and its Henry coefficient $S_i$:
\begin{equation}
    P_i = D_i \cdot S_i.
\end{equation}
As a result, the molar flux of component i through the film $\dot{n}_i^{\prime \prime}$ is described by:
\begin{equation}
    \dot{n}_i^{\prime \prime}=\frac{D_i \cdot S_i}{\delta} \cdot\left(p_{feed,i} - p_{permeate,i}\right),
\end{equation}
where $p_{feed,i} $ and $ p_{permeate,i}$ are the partial pressures of component \textit{i} in the feed and permeate compartment. On the one hand, the diffusivity of a gas molecule is strongly dependent on its size, where smaller molecules permeate faster. On the other hand, the Henry coefficient depends on the gas molecule's solubility in the polymer matrix. Dependent on the properties of the polymer, either the diffusivity or the solubility is the rate-limiting step and controls the membrane's selectivity.\cite{baker2012membrane}

\subsection{Modeling transport processes through coated films and membranes} \label{subsec22}

On the microscopic scale, the interactions of permeating molecules and the pore wall dominate mass transport for mechanisms such as capillary condensation, surface diffusion and configurational diffusion and cannot be neglected for Knudsen diffusion. Therefore, molecular dynamics (MD) simulations proved an effective approach for modeling permeation in this domain. On a macroscopic scale, pore wall interactions exert only small influence on mass transport rates and can therefore be neglected. This enables the consideration of the entire system consisting of polymer substrate and thin film coating, compared to the MD approach which only considers a small fraction of the system, a.o. due to limited computational capacities.

An early approach for macroscopic modeling of the permeation was made by Rossi and Nulman \cite{rossi1993}, who established a numerical model of permeation through defects in a barrier layer. This was further analyzed and combined with defect detection methods by others \cite{dasilva2000,hanika2003}. This thread was picked up by Wilski \textit{et al.} \cite{Wilski_2017}, who took advantage of enhanced computational capacities by focusing on smaller defect sizes. They established a model for simulating the oxygen transfer rate (OTR) through a polyethylene terephthalate (PET) substrate with a thin film barrier coating containing pores on a macroscopic level. They considered defects in the range of 0.3 µm – 0.7 µm in the coating and assumed the pristine coating to be impermeable while in the defect area, ambient oxygen concentration was present on the PET surface. For PET, solution diffusion was assumed, and the oxygen transport was numerically calculated by solving Fick’s Law. This allowed for the investigation of the effect of adjacent defects on mass transport and a critical defect distance was found, under which interaction occurred which reduces the mass transport rate per defect.

In a following paper, Wilski \textit{et al.} \cite{Wilski_2021} compared this simulation to a composite of a stainless-steel film bonded to a PET film. Before the bonding, the steel films were perforated with a set of laser drillings with a defined diameter and spacing, to approximate the macroscopic porosity of a thin film barrier coating while eliminating the influence of sub-macroscopic pores and defects on mass transport. They found similar behavior in the change of OTR as a function of defect spacing, but the simulation underestimated the critical defect distance and the barrier performance. This gap was ascribed to the experimental setup, where the artificial defects created by laser drilling differed from the model’s defect. Due to the drilling process, it was not possible to create cylindrical pores and additionally, the adhesive used for the bonding influenced the oxygen concentration distribution. This resulted in an effective pore diameter that could not be determined exactly. Nonetheless, the macroscopic approach has made a valuable contribution to the understanding of permeation through thin film barrier coatings on polymer substrates, as they enabled the consideration of the entire system. In a next step, the combination with the microscopic approach is needed to create a multiscale model that considers the permeation mechanisms in an exhaustive manner.\\

\subsection{Modeling transport processes at the molecular scale} \label{subsec23}

As already alluded to in the previous subsection, MD simulations are particularly appropriate to model the structure and dynamics of molecules in porous materials and as such the permeation process on at atomistic level. More precisely, MD involves the numerical solution of Newton’s equation of motion, thus allowing to compute both equilibrium thermodynamic and dynamical properties at realistic temperature and pressure conditions. As such, MD simulations can be thought as making the atomic real-time evolution of the atoms visible, which is why another role of MD is that of a computational microscope \cite{Kuehne2014}.

One of the most important, but at the same time computationally demanding, factors is the calculation of the interatomic interactions between the atoms for every timestep during an MD simulation. In conventional, so-called classical MD simulations, the interatomic forces are determined as the nuclear derivatives from \textit{a priori} parametrized empirical potential functions. So far they are mostly parametrized so as to reproduce experimental measurements \cite{Kuehne2015}, though the current state-of-the-art are so-called neural-network or machine learned potentials to mimic accurate \textit{ab-initio} data \cite{Ghasemi2022}. Yet, in spite of great progress, they are typically not able to simulate chemical bonding processes that take place in the interior of pores with sufficient accuracy and the transferability to systems or regions of the phase diagram that are different to ones of the assumed in the fitting process is often restricted.

Hence, a first-principles based simulations, such as ab-initio molecular dynamics (AIMD) \cite{Hutter2023}, where the forces are computed on-the-fly from accurate quantum mechanical calculations, is a very attractive alternative since many of these limitations can in principle be removed, even though the increased accuracy and predictive power of AIMD simulations comes at significant computational cost. For that purpose, beside the well established Born-Oppenheimer and Car-Parrinello AIMD methods \cite{Car1985}, a third and much more efficient computational technique had been developed that aims at unifiying the best of both approaches \cite{Kuehne2007}. This is nowadays known as second-generation Car-Parrinello AIMD and it is the leading scheme for complex large-scale applications such as porous systems \cite{Prodan2018,Prodan2020}.

Interestingly, however, beside the choice of computational simulation method, the generation of an atomistic pore model that is as small as possible, but at the same time accurately represents the essential chemistry of physics, is rather subtle and challenging. A rather promising and robust protocol that has emerged of the present research is to first create a liquid silica melt by heating crystalline $\beta$-cristobalite silica to well above 3000~K. Amorphous bulk silica can then be obtained by quenching it from the melt via a dynamical simulated annealing procedure \cite{Camellone2009}.
Other approaches use ReaxFF simulations with similar annealing procedures~\cite{Allolio_2014}. At this stage, the system's density is 2.31 g/cm$^{3}$, which is very close to P. Gallo’s MCM-41 model~\cite{Gallo2010} and between the densities of quartz (2.6 g/cm$^{3}$), as well as fused silica (2.2 g/cm$^{3}$). Within experimental works of others an apparent density of 2.37 g/cm$^{3}$ had been reported for MCM-41 silica~\cite{Weinberger2016}.

The desired silica pores can then be created by removing a cylindrical volume with a specific predetermined diameter from the amorphous structure.

For hydrophilic systems the free valences on the inner surfaces were saturated via additional hydroxyl groups, whereas for hydrophobic systems, several of the hydroxyl groups were exchanged by trifluoromethyl groups (CF$_{3}$). This procedure entails a relatively equal distribution of the groups on the surface with functional group densities varying between 1.75 and 2.96 nm$^{2}$, which corresponds to 23.8\% to 42.8\% of functional groups being CF$_{3}$ within hydrophobic pores.

Since filling the whole pore volume with water would lead to an unrealistically large water density and potentially deform the pore walls, many computational studies use simple cylindrical pores with a defined pore size and fill the the pore volume with enough water molecules to reach the experimental bulk water density~\cite{Allolio_2014,hartnig2000}. Alternatively, nanopores can also be created by employing the volume-scaling method to set the density to the experimentally observed value for a given composition~\cite{Nakano1994,Kieffer1988}. In this way, simulated and experimental porosities were found to agree well with each other for amorphous silica, hydrogenated amorphous silica, and nanoporous hydrogenated amorphous silica~\cite{Gergs2022}. Another way to address this issue is to perform MD simulations in the isothermal-isobaric ensemble~\cite{mosaddeghi2012, ghoufi2011}, or via the Monte Carlo method in the grand-canonical ensemble, i.e. with variable particle numbers~\cite{shroll1999molecular,shevade2000molecular}. But, this is only possible when using classical MD since the necessary timescales renders it unfeasible for AIMD simulations.


This is why in most previous simulation studies conventional force-field based MD simulations were done~\cite{Lerbret_2011,Bourg_2012,Renou_2014,Gallo_2012}. In only very few studies density functional theory-based AIMD MD simulations have been conducted~\cite{Sulpizi_2012,Cimas_2014,Allolio_2014}, though only for rather small and simplistic model systems. For that reason, the work of Weinberger \textit{et al.}~\cite{Weinberger_2022}, which reported on quantum mechanical semi-empirical MD simulations that had been optimized for water \cite{Voorhis2015}, is particularly noteworthy and allowed to calculate the translational diffusion constant as a function of pore size and hydropathy for a large variety of different systems.


\section{Analysis of micropores in thin plasma polymer films}\label{sec3}

\subsection{Experimental analysis}\label{subsec31}


The analysis of micropores in plasma polymer films is of utmost importance for the understanding of barrier and membrane properties as well as corrosion protection properties. However, only few techniques provide insight in the nanometer and even sub-nanometer small pores that are characteristic for plasma deposited amorphous films. While surface microscopic tools such as AFM and electron microscopy mainly allow for the analysis of macroscopic defects and do not provide insight in the micropore structure, transmission electron microscopy offers the possibility to analyze nanoscopic defects \cite{Kuo.2015}. However, based on TEM only small areas of the coating can be analyzed and correlation to barrier or membrane properties and the application of TEM has been mainly limited to the analysis of metal/plasma polymer nanocomposite films \cite{Alissawi.2013}.

A well-known method for the analysis of the microstructure of thin plasma polymer films is ellipsometric porosimetry (EP). The method is based on the change of the optical properties of thin films during adsorption or desorption of a volatile species either at atmospheric or under reduced pressure \cite{Perrotta.2015}. A particularly powerful development is the combination of EP with electrochemical impedance spectroscopy (EIS) for the analysis of the films \cite{Perrotta.2015}. EIS is based on the analysis of the frequency dependent complex impedance of the plasma polymer applied to an electrically conducting substrate in contact with an ion-conducting electrolyte \cite{Perrotta.2015,Perrotta.2015b}. Perrotta \textit{et al.} successfully applied this combination for the characterization of the nano-porosity and water permeation properties of plasma deposited SiO$_x$ layers. They could also show that dynamic EP also allows the detection of macroscale defects  using probe molecules not penetrating the bulk nano porosity of the layer \cite{Perrotta.2016}.

EIS can also be combined with other optical characterization methods. Recently, Xie \textit{et al.} combined the EIS analysis with the in-situ IR spectroscopic analysis of water in HMDSN and HMDSO based thin plasma polymer films \cite{Xie.2022}.

For ultra-thin films containing open pores and for the analysis of initial stages of film formation the application of cyclic voltammetry offers the possibility to analyze non-covered substrate areas quantitatively \cite{Liu_2003,Ohto_2015,Hoppe.2017}. In this case, the peak current densities of a redox mediator depend on the actual electrode area in contact with the electrolyte. By comparing bare and thin film covered electrode surfaces, a quantification of the free surface sites becomes possible. Oezkaya \textit{et al.}, Liu \textit{et al.} and Hoppe \textit{et al.} could characterize the defect density in ultra-thin films as a function of the substrate chemistry based on such an approach \cite{Liu_2003,Ohto_2015,Hoppe.2017}.

For the quantitative analysis of the pore size dimensions, Hoppe \textit{et al.} used positron annihilation lifetime spectroscopy (PALS) and combined this approach with diffuse reflection Fourier transform infrared spectroscopy (DRIFTS) of SiO$_x$ thin films \cite{Hoppe.2020}. The combination of these two techniques has proven to be particularly fruitful. On the one hand, PALS enables the analysis of the electron density distribution in a wide variety of materials and is thus an established technique for the non-destructive characterisation of atomic open defects (vacancies, clusters, etc.) \cite{Hoppe.2022}. It can be applied in solids and can be used to determine the free volume and pore size distribution in amorphous and porous media. On the other hand, the application of IR spectroscopy to the determination of internal structure and cross linking in SiO$_x$ materials has been widely establish, as recently reviewed by de los Arcos \textit{et al}. \cite{delosArcos.2021}. Through the combination of PALS and IR spectroscopy, Hoppe \textit{et al.} showed a clear correlation between pore size distributions in the plasma-deposited films and the SiO$_x$-peak position in the infrared spectra \cite{Hoppe.2020,Hoppe.2022}.

\subsection{Interpretation of porosity data}\label{subsec32}


Currently, no analytical method is available for porosity analysis of a coating that is capable of capturing the entire spectrum of possible pore sizes. For further interpretation of porosity measurements, the different dimensions in which the methods can be restricted in their scope should be carved out.

On a micro- and nanoscale level, a thin film consists of areas filled with its backbone matter e.g., SiO$_x$, and void areas, i.e. pores or defects. Mostly, they are considered cylindrical or spherical, depending on the measurement, with the diameter as a characteristic parameter, often referred to as “pore size”. The pore size distribution is obtained by creating a histogram of the pore sizes. However, this does not consider important aspects of the porous structure. Clusters of interconnected pores can impose significantly different effects on the thin film permeability than a similar volume of single, closed pores. This is especially true for pores or pore networks that are located at one or both interfaces of the thin film (“open porosity”). They offer permeation paths to gases and thus significantly influence the barrier or membrane properties. The permeability of these paths depends on their cross-sectional area and tortuosity (ratio of path length to film thickness \cite{Menzel.2001}. As most of these do not connect both interfaces of the thin film, their relative position to other pores/ pore clusters is another important factor as it determines the path length for solution diffusion. Thus, the relevance of the 3D arrangement of a thin film’s pores should always be considered when analyzing and comparing porosity data.

\begin{figure}[h]
\centering
\includegraphics[width=0.48\textwidth]{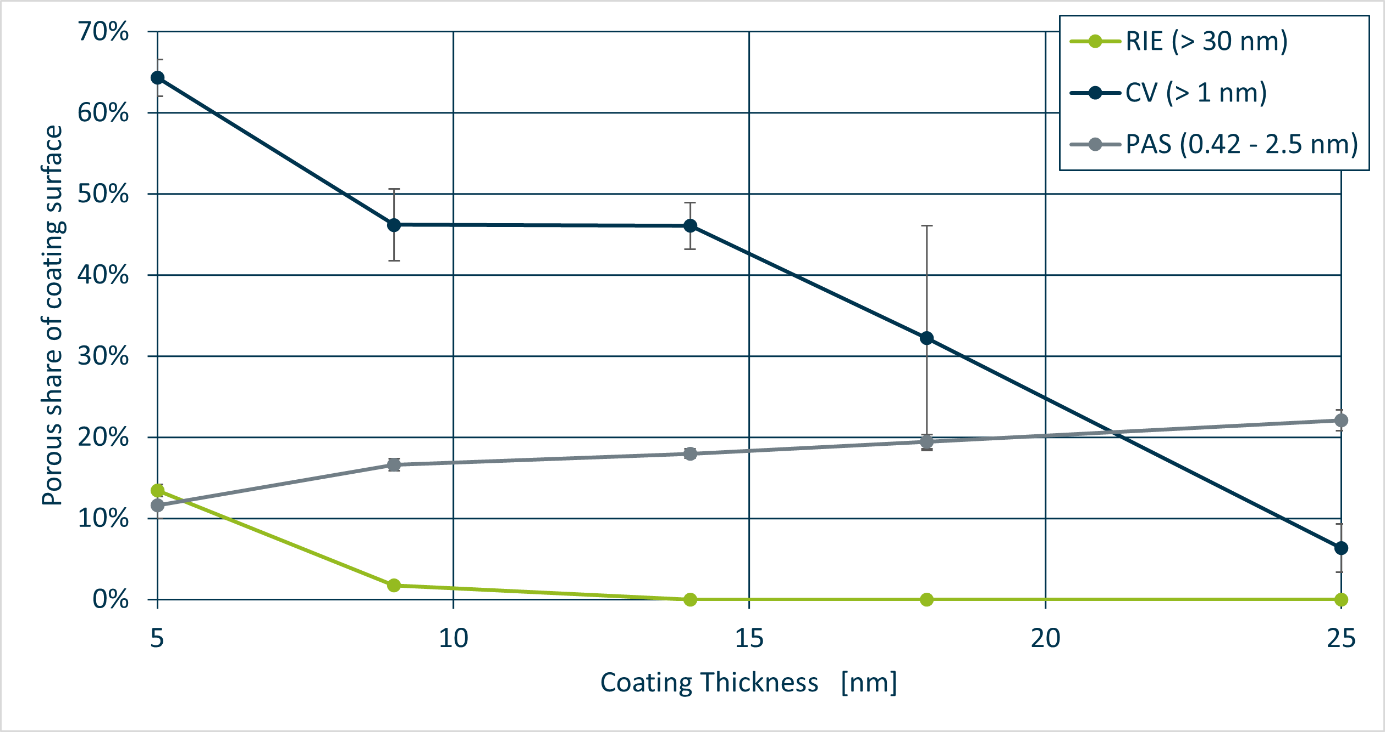}
\caption{Porosity percentage of total area according to different methods as a function of coating thickness. Pores larger than 30 nm are only present to a small extend and quickly vanish with increasing coating thickness. The area covered with pores larger than 1 nm also declines with coating thickness, but is significantly larger especially at low coating thicknesses. Smaller pores seem to slightly increase with coating thickness. Data based on the work of Wilski \textit{et al.} \cite{Wilski_2020} and \cite{Wilski.2022}.}
\label{fig:IKV1}
\end{figure}

The interpretation of a combination of different porosity measurement methods can yield valuable information. Wilski \textit{et al.} \cite{Wilski_2020} combined reactive ion etching (RIE) and cyclovoltammetry (CV) to analyze SiO$_x$ coatings on PET with different coating thicknesses and compared it to OTR and water vapor transfer rates (WVTR). PALS data of the same samples was amended, which will be published in more detail elsewhere. All porosity measurements consider different aspects of the porous structure: With RIE, only open porosity > 30 nm can be observed, but it is possible to create a pore size distribution. CV is able to detect open porosity > 1 nm, but only provides the share of surface area covered by these pores. Finally, the PALS creates a depth profile of defects in the range of 0.42 nm – 2 nm, where it determines an average defect size on distinct implantation depths. It also provides a relative intensity of the measurement signal, which does not equal the volumetric porosity but can be considered proportional. In order to compare the different results of porosity measurements, the porosities are projected on a 2D surface. For this, the depth profile of the relative intensities of PAS are averaged out and are interpreted as a porosity projected onto a 2D area. This represents a simplification and PALS data should be considered with a caveat. The resulting comparison is shown in Figure \ref{fig:IKV1}. These data can be used to ascribe different diffusion mechanisms according to the pore size, as determined by Wilski \textit{et al.} \cite{Wilski_2020} (see also Subsection \ref{subsec21}). In addition, this enables the correlation with permeation data to identify dominant transport mechanisms through thin film coatings. Further investigations should focus on these correlations while generating a more comprehensive representation of the porosity.

\section{Tuning of thin film barrier properties and permeability}\label{sec4}


\subsection{Barrier properties}\label{subsec41}

\begin{figure}[h]
\centering
\includegraphics[width=0.48\textwidth]{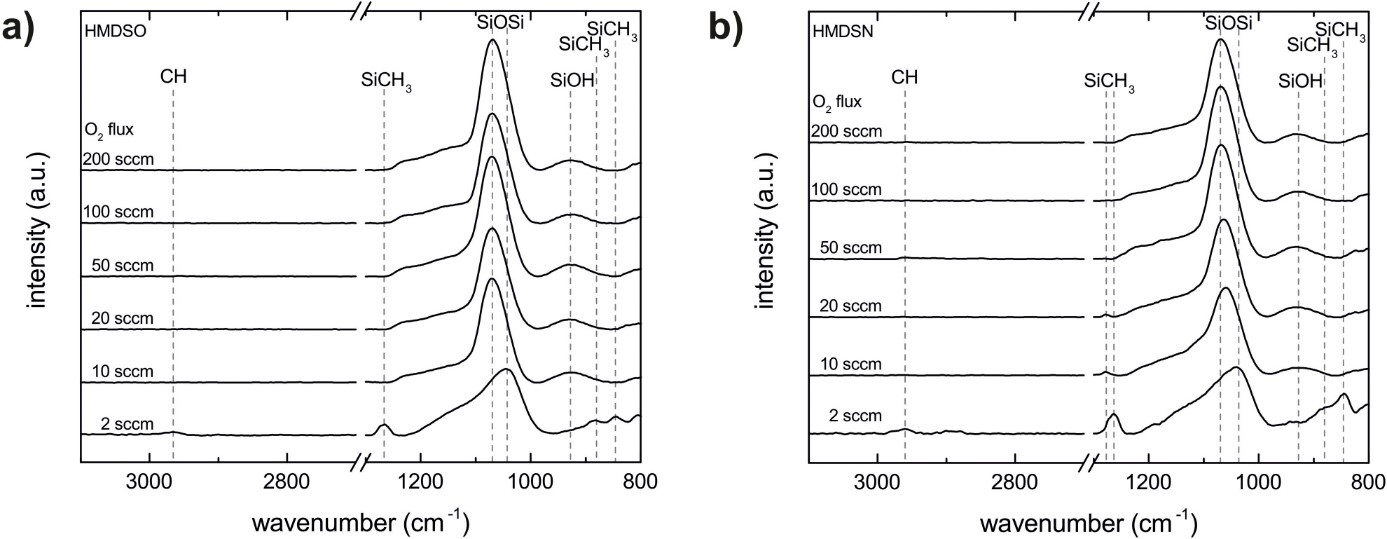}
\caption{FTIR spectra of \textbf{(a)} HMDSO and \textbf{(b)} HMDSN barrier coatings as a function of oxygen admixture; the shift of the Si-O-Si peak towards higher wavenumbers indicates higher crosslinking density. Reprinted with permission from IOP Publishing \cite{Mitschker.2018}.}
\label{fig:IKV2}
\end{figure}

Properties of silicon based PECVD coatings can be tailored to different applications by carefully adjusting the deposition parameters e.g., precursor and gas fluxes, induced energy, process pressure, and temperature. For silicon organic precursors, such as Hexamethyldisiloxane (HMDSO) or hexamethyldisilazane (HMDSN), for example, the ratio between precursor and oxygen as well as the induced power is crucial for the structure and chemistry of the deposited thin films. Higher amounts of oxygen relative to the precursor result in more inorganic thin films with less carbon and hydrogen incorporated \cite{Bahroun_2014}. Depending on the power input in the microwave (MW) driven PECVD process the energy density in the plasma can be controlled. Higher energy density, induced through higher power input, increases the fragmentation of the precursor and therefore yield higher crosslinking density. Vice versa, lower oxygen flow and energy density results in lower fragmentation and more incorporation of carbon and hydrogen resulting in lower crosslinking density and more organic SiOCH thin films \cite{Mitschker.2018}. Maintaining a surplus of oxygen within the deposition process results in shifting of the Si-O-Si bond towards higher crosslinked structures due to increased oxidation of the already deposited thin film (Figure\ref{fig:IKV2}) \cite{Mitschker.2018,Jaritz.2017}.

\begin{figure}[h]
\centering
\includegraphics[width=0.48\textwidth]{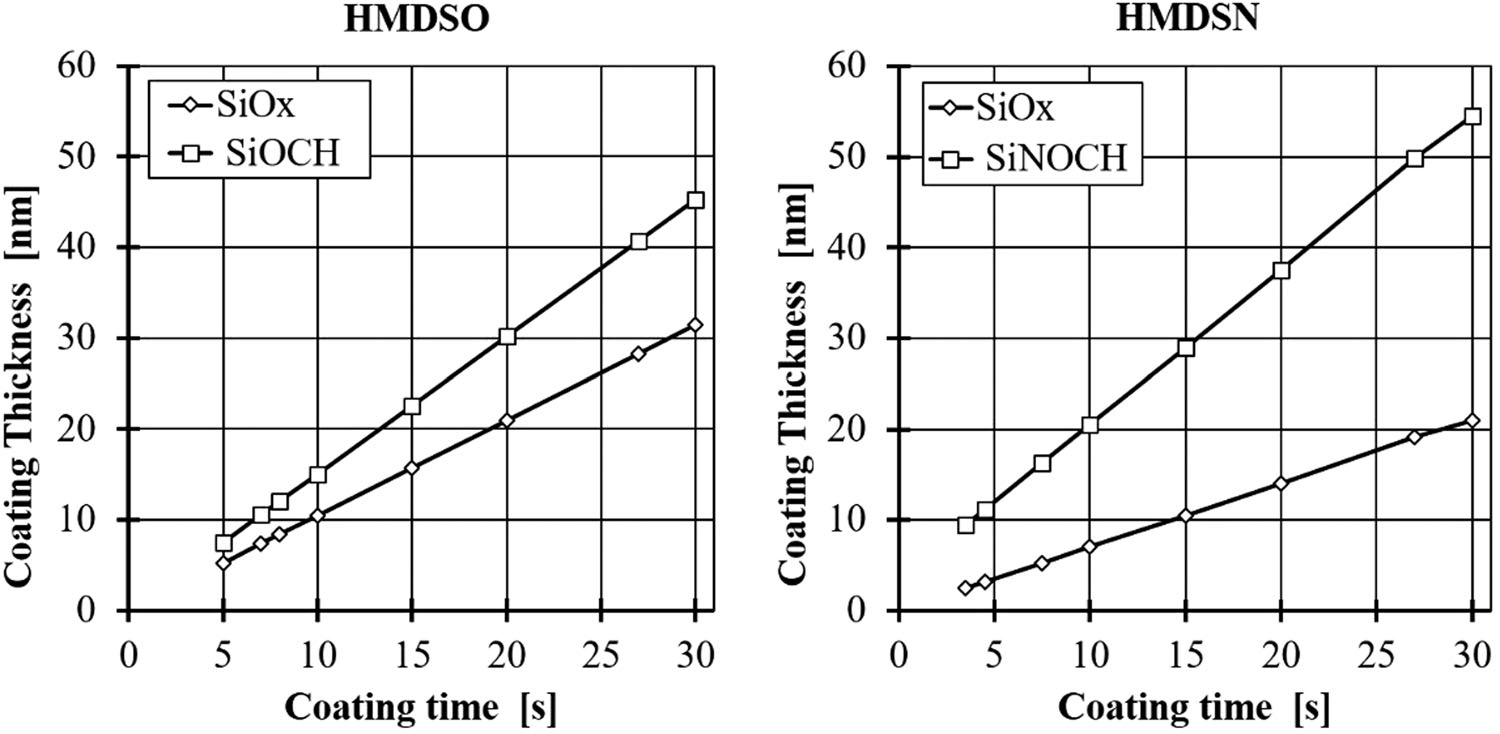}
\caption{Coating thickness over time for silicon based thin films deposited from HMDSO and HMDSN as precursors. Reprinted with permission from Wiley \& Sons  \cite{jaritz2021}.}
\label{fig:IKV3}
\end{figure}

Compared to HMDSO, HMDSN-fed processes for SiO$_x$ barrier layers result in higher barrier performance under similar deposition conditions in regard to precursor-to-oxygen ratio, pressure, peak MW-power and duty cycle. Under the same deposition conditions (i.e., power input, coating time and gas flow) HMDSN yield a lower deposition rate compared to HMDSO (Figure \ref{fig:IKV3}). Since binding energies within the HMDSN molecule are lower compared to HMDSO less energy per molecule is needed for fragmentation. This leads to higher fragmentation in HMDSN fed processes and denser SiO$_x$ thin films \cite{Mitschker.2018,jaritz2021}.
A dense and coherent thin film is the basis for good barrier performance against permeation of e.g., oxygen or carbon dioxide. Therefore, the critical thickness of the thin film is reached when the whole substrate surface is covered with a coherent thin film. Regardless of the used monomer the critical thickness is approximately 5 nm \cite{jaritz2021}.

The substrates surface morphology effects the structure of the deposited barrier thin film. Therefore, the application of silicon-organic interlayers is an approach to enhance the performance of barrier thin films. For example, anti-block particles in the PET substrate film interfere with the growth of the SiO$_x$ barrier thin films which results in defects and reduced barrier performance. Applying a silicon organic interlayer improves the coherent growth of the subsequently deposited SiO$_x$ barrier thin film by providing a smoother surface. For the deposition of organosilicon interlayers, however, higher fragmentation of the HMDSN molecule favors volume polymerization within the plasma and cause higher deposition rates of SiNOCH films compared to HMDSO based SiOCH interlayers (Figure \ref{fig:IKV3}). With the higher deposition rates of SiNOCH interlayers the structure of those is much more granular in contrast to SiOCH interlayers. Thus, SiNOCH interlayers do not benefit the growth of SiO$_x$ barrier thin films onto PET \cite{jaritz2021}.
However, deposition of SiOCH interlayers in a microwave driven PECVD process onto PET can result in a granular structure of the interlayer which propagates into the barrier thin film and reduces the barrier performance, while interlayers deposited in a capacitively coupled plasma (CCP) does in fact improve the growth of the subsequent SiO$_x$ barrier layer due to compensation of surface imperfections \cite{Bahroun_2014}. Therefore, careful adjustment of the process parameters, i.e., oxygen/monomer ratio, duty cycle, power input, depending on the substrate is necessary to ensure a good barrier performance.

\textbf{Adhesion of barrier thin films on polyolefin surfaces:} Due to their low surface free energy, coatings on polyolefins are generally subject to the problematics of low adhesion. For this reason, polyolefins must be suitably pretreated. This can be achieved by applying a silicon organic interlayer which does not have significant barrier properties but interconnects the polymer and the barrier layer. Behm \textit{et al} showed that the pre-treatment of PP enhances the adhesion of SiOCH interlayers by approximately two times compared to untreated surfaces. Regardless of whether oxygen or argon is used for pre-treatment, comparable improvement can be achieved if the ion-fluence is adjusted accordingly \cite{Behm.2014}. However, PP is prone to overtreatment which results in the formation of an oxidized polymer layer with low molecular weight and reduced cohesion. Therefore, this weak boundary layer compromises the adhesion of subsequent applied thin films. Jaritz \textit{et al} found that UV radiation emitted by oxygen plasma plays a crucial role in adhesion of SiOCH coatings deposited on PP \cite{Jaritz.2017b}.  Oxygen plasmas emit UV radiation with clear peak intensities at 130 nm and 260 nm which is able to break the C-C backbone structure of PP and PE respectively. These reactive scissions are points of bonding with the SiOCH interlayer. While for oxygen plasmas a clear contribution of the UV radiation was observed, since UV treatment solely leads to the same adhesion enhancement compared to plasma pretreatment when the treatment time is adjusted, in case of argon plasmas ion bombardment was identified as the main factor for successful pretreatment \cite{Jaritz.2017b}.

\begin{figure}[h]
\centering
\includegraphics[width=0.48\textwidth]{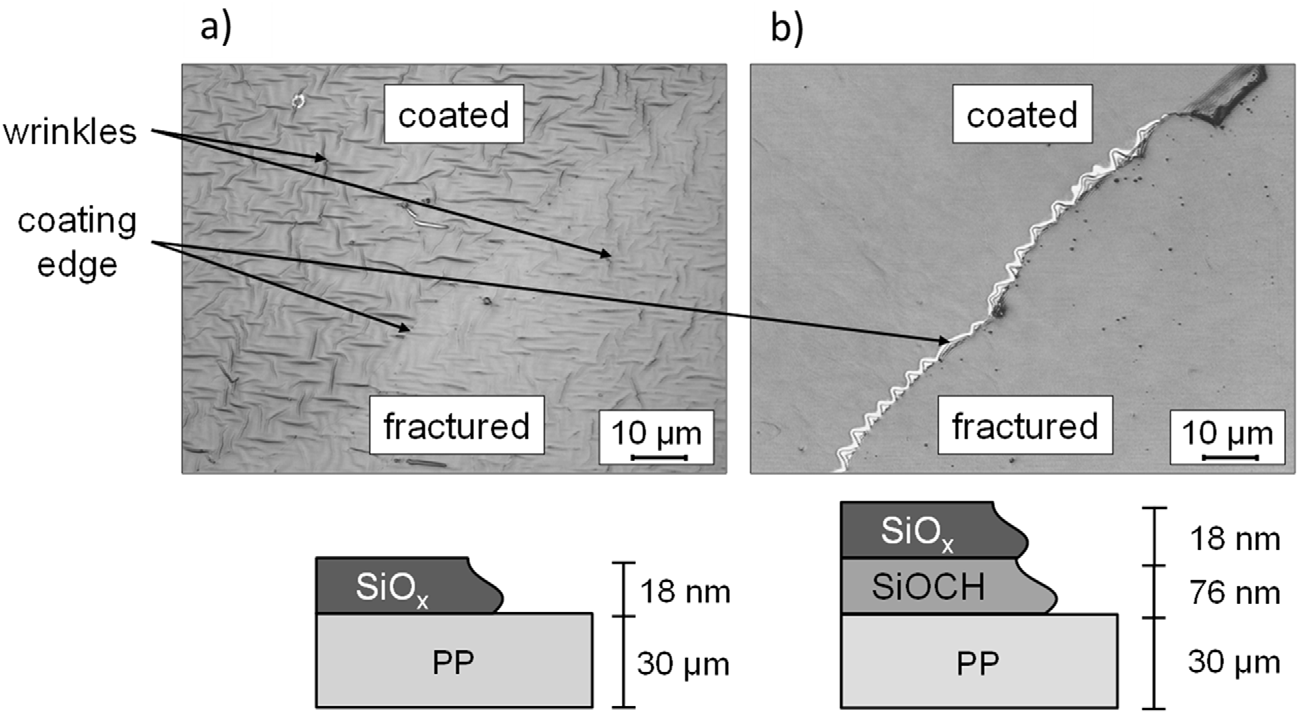}
\caption{Wrinkle formation of SiO$_x$ coated PP \textbf{(a)} and effect of a SiOCH interlayer \textbf{(b)}. Reprinted with permission from IOP Publishing \cite{Jaritz.2017}.}
\label{fig:IKV4}
\end{figure}

\textbf{Stresses in barrier thin films:} The brittle nature of SiO$_x$ can cause failure of the thin film due to internal and external stresses. Therefore, residual stresses are relevant for any application. Applying SiO$_x$ thin films directly onto the PP substrate results in formation of so-called wrinkles probably caused by the heating of the surface during the high power deposition process (Figure \ref{fig:IKV4} ). For the deposition of SiOCH interlayers considerably less power is needed for deposition, thus, the formation of wrinkles can be avoided. The application of a SiOCH interlayer also provides protection of the PP surface against over-oxidation by the surplus of reactive oxygen species during the SiO$_x$ deposition process \cite{Jaritz.2017}.

At a constant thickness of approx. 35 nm, SiO$_x$ thin films show increasing compressive stresses with longer pulse-on times in pulsed microwave plasma excitation. Investigation of the structure shows higher crosslinking densities as well as more inorganic character, due to higher oxygen incorporation, of coatings deposited at longer pulse-on times. Most of the thin film deposition processes take place in the first 2 ms of the pulse, while oxidation and crosslinking processes dominate afterwards \cite{Jaritz.2017}.

\begin{figure}[h]
\centering
\includegraphics[width=0.48\textwidth]{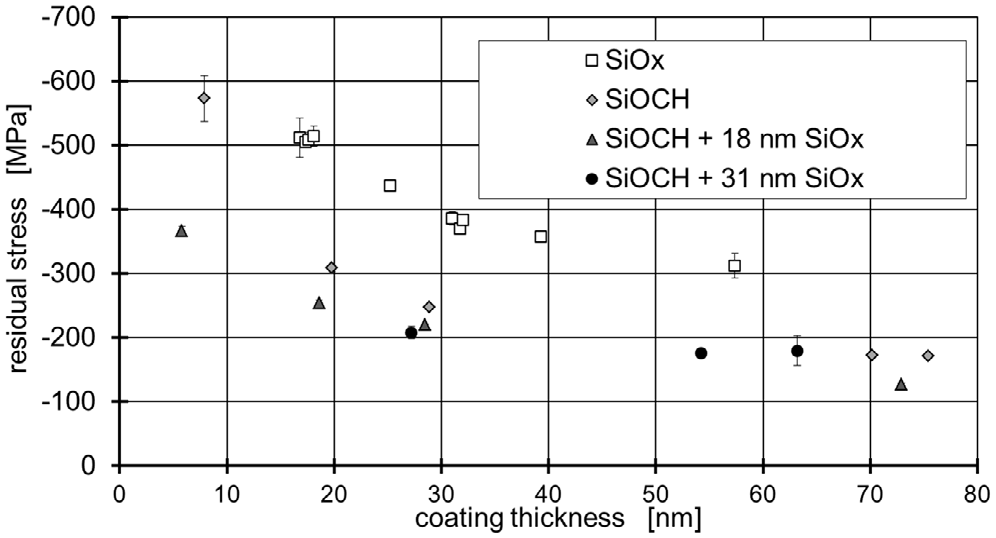}
\caption{Residual stresses of SiO$_x$ and SiOCH coatings and different SiO$_x$ barrier thickness Reprinted with permission from IOP Publishing \cite{Jaritz.2017}.}
\label{fig:IKV5}
\end{figure}

The thickness of the thin film is crucial for the resulting stresses. Depending on the thickness of the SiOCH interlayer as well as the subsequent deposited SiO$_x$ barrier thin film residual stresses could be reduced by the factor of approx. 3.5 (Figure \ref{fig:IKV5}) \cite{Jaritz.2017}.

\begin{figure}[h]
\centering
\includegraphics[width=0.48\textwidth]{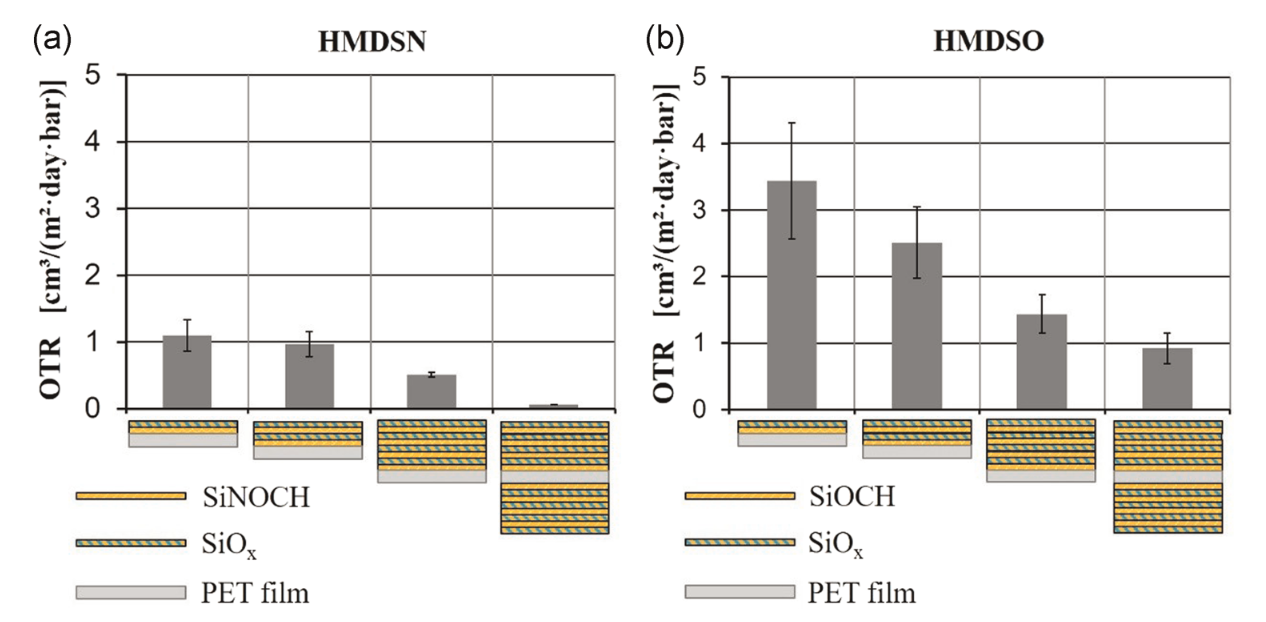}
\caption{Residual stresses of SiO$_x$ and SiOCH coatings and different SiO$_x$ barrier thickness. Reprinted with permission from Wiley \& Sons \cite{jaritz2021}. }
\label{fig:IKV6}
\end{figure}

\textbf{Multilayer Systems:} An improvement of barrier performance regarding oxygen permeation can be achieved by deposition of alternating silicon-organic interlayers and SiO$_x$ barrier layers, so called dyads (Figure \ref{fig:IKV6}). Generally multilayer systems deposited from HMDSN result in better barrier performance regarding oxygen transmission than multilayer systems from HMDSO as a monomer. This corresponds with the denser film growth of SiO$_x$ barrier thin films as discussed before. Since the barrier performance of SiO$_x$ thin films is defect driven i.e., depending on the distribution and size of defects within the thin film, alternating silicon-organic  and SiO$_x$ lengthen the diffusion path though the coating system according to Nielsens tortous path model.  This model explains the better barrier performance by stating that the disconnection of films with randomly placed defects, in case of barrier thin films pores, by stacking  intermediate layer with no barrier function results in misalignment of defects thus a higher barrier performance due to hindered permeation.  The thickness of the intermediate layer, however, should not be greater than 6 nm for the reason that thicker intermediate layers grow more granular which contradicts a coherent SiO$_x$ thin film growth \cite{jaritz2021}.

\subsection{Membrane properties }\label{subsubsec42}

\subsubsection{Experimental studies}
In contrast to barrier coatings, membrane films have to allow certain species to permeate through them. Various studies have been carried out to gain knowledge on the impact of different fabrication parameters on the permeation behavior of the deposited films.
For the investigation of the permeation behavior of the deposited films, a thin film composite membrane with a about 160~nm polydimethylsiloxane layer as top layer produced by Helhmholtz Zentrum Hereon was used \cite{yave2010nanometric}. All coatings were deposited in a low-pressure plasme reactor by plasma-enhanced chemical vapor deposition. Four microwave magnetrons generate pulsed microwaves with a frequency of 2.45 GHz. Detailed information about the reactor design can be found elsewhere \cite{jaritz2020hmdso}.

The first study, conducted by Kleines \textit{et al.}, used HMDSN and investigated the influence of auxiliary gases (argon and nitrogen) on the chemical and membrane properties of the deposited film \cite{kleines2020enhancing}. Pure HMDSN and HMDSN with argon or nitrogen with a ratio of 2:5 were used at constant fabrication parameters (see \cite{kleines2020enhancing} for detailed information).
AFM measurements have shown that the auxiliary gas has a significant influence on the morphological structure of the deposited layers. The coating chemistry analysis showed a decrease in silicon carbide content in the films fabricated with argon to none auxiliary gas to nitrogen. At the same time, the oxygen content of the layer deposited with nitrogen is significantly higher than in the other two layers after post-oxidation. The decrease of silicon carbides and the increase of oxygen incorporated in the coatings is accompanied by an improvement in permeability. This indicates that increasing the number of flexible siloxane bonds while simultaneously decreasing rigid silicon carbide bonds, leads to improved intrasegmental mobility. The resulting selectivities clearly overcome Knudsen selectivity and are best for nitrogen as auxiliary gas. \cite{kleines2020enhancing}

In contrast to the first membrane study, all subsequent PECVD fabrications used HMDSO as precursor. Kleines \textit{et al.} investigated the influence of different microwave peak powers as well as oxygen to HMDSO ratios on the chemical structure and membrane performance. The detailed fabrication parameters can be found elsewhere. \cite{kleines2021structure}

\begin{figure}[h]
\centering
\includegraphics[width=0.33\textwidth]{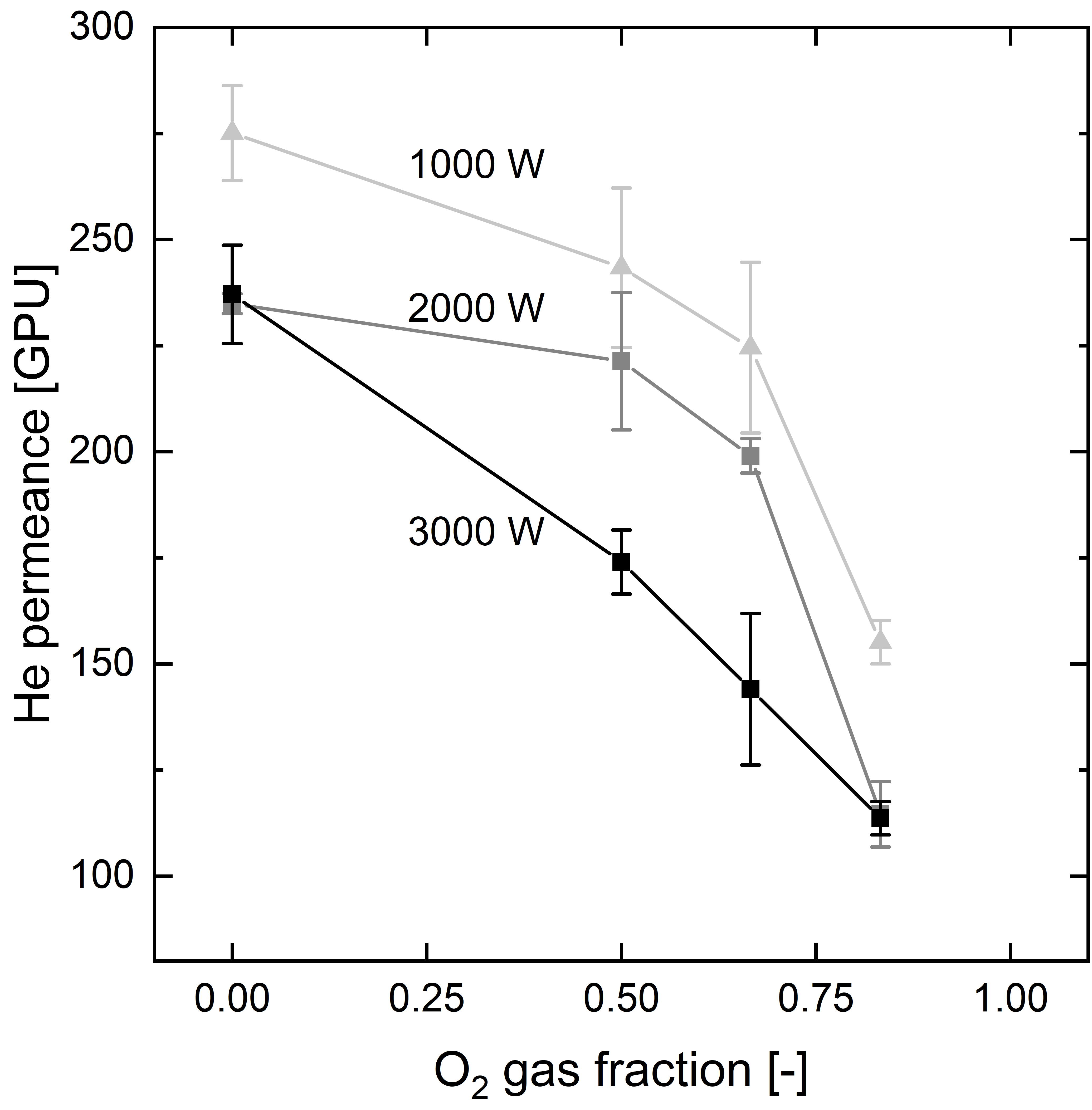}
\caption{He permeance of organosilica membranes fabricated with microwave power inputs of 1000, 2000, and 3000~W with \ce{O2} gas fractions in HMDSO from 0 to 0.83. Each data point consists of the measurements of three membrane samples.}
\label{fig:P_vs_O2 - He}
\end{figure}

Figure \ref{fig:P_vs_O2 - He} shows the helium permeance of the deposited organosilica layers for the investigated peak powers and \ce{O2} to HMDSO ratios. Since helium is the smallest inspected gas molecule and the parameters were optimized for the best size-sieving separation, we strived for a high helium permeance. Low microwave peak powers together with low \ce{O2} content lead to the highest permeances.\\

\begin{figure}[h]
\centering
\includegraphics[width=0.33\textwidth]{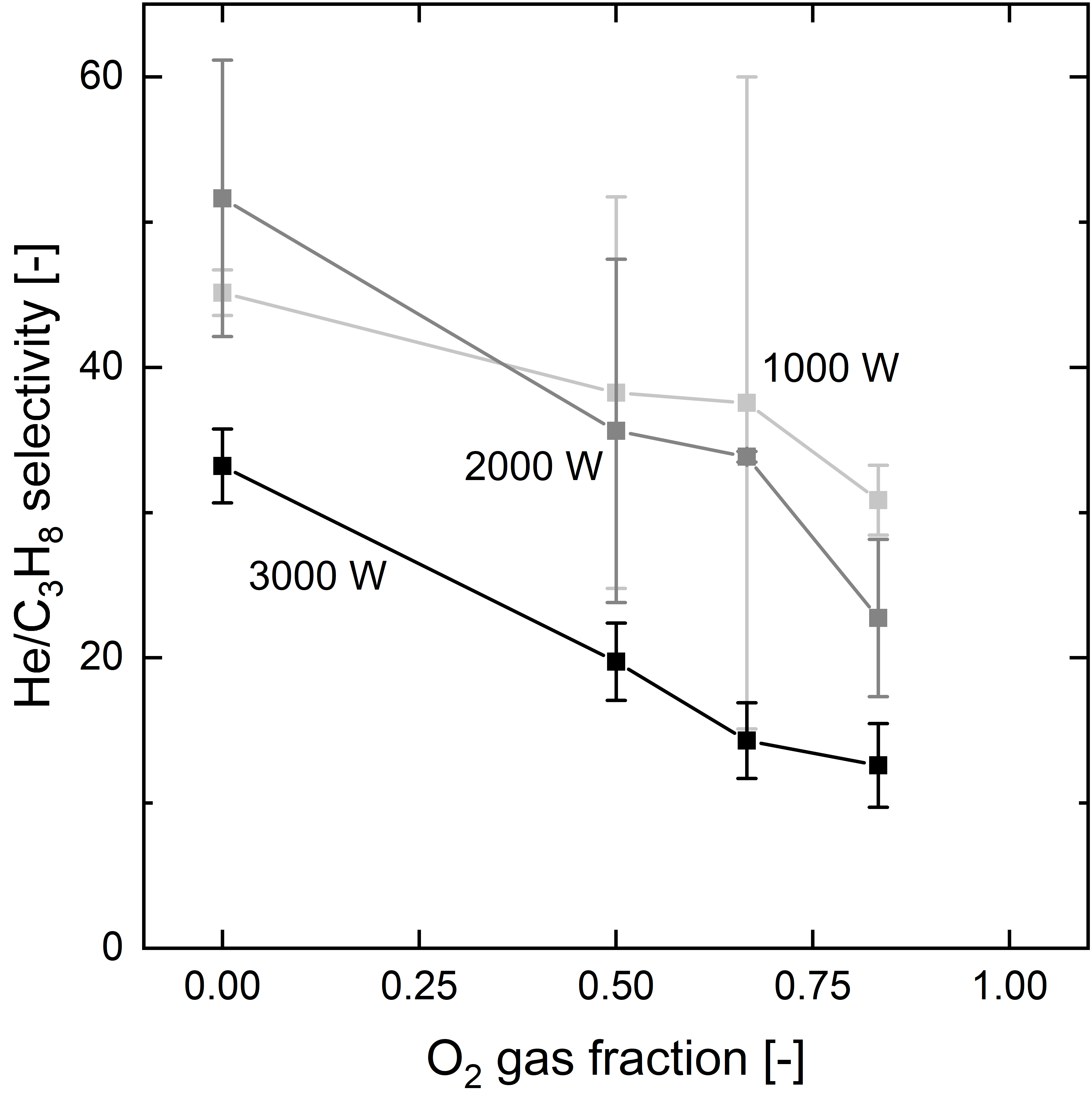}
\caption{He/\ce{C3H8} selectivity of organosilica membranes fabricated with microwave power inputs of 1000, 2000, and 3000~W with \ce{O2} gas fractions in HMDSO from 0 to 0.83. Each data point consists of the measurements of three membrane samples.}
\label{fig:S_vs_O2 - HeC3H8}
\end{figure}

To evaluate the separation performance, also permeances for \ce{CO$_2$, N$_2$, CH$_4$, and C3H8} were measured \cite{kleines2021structure}. Figure \ref{fig:S_vs_O2 - HeC3H8} shows the resulting selectivity for He to \ce{C3H8}. Since those two molecules have the biggest difference in molecule size of the investigated gases, the trends are most significant. Here again, the selectivities are highest for the organosilica layers deposited at low microwave peak powers and least oxygen in the plasma feed gas.\\

Regarding their chemical structure investigated via X-ray photoelectron spectroscopy (XPS) and Fourier transform infrared spectroscopy (FTIR), the best performing membranes show a low atomic oxygen content and a high number of methyl groups. Hence, the separation behavior profits from a more organic/less oxidized chemical structure.

Those findings are also in agreement with the ones published here \cite{kleines2022evaluation}. There, Kleines \textit{et al.} investigated the influence of the pulse-on time on the chemical structure and membrane properties of the deposited layer \cite{kleines2022evaluation}.
Organosilica films were deposited in a pulsed microwave excited low-pressure PECVD process with a thickness of about 100~nm. The film depositions took place under low energy input in the energy-deficient range. The dependence of the layer and separation properties on the MW power input and pulse duration were investigated as a function of the extended energy density factor (Yasuda factor).
It systematizes the properties and categorizes properties that correlate with the plasma process settings. However, the measured properties for the layer systems show that differences in the properties, in particular, due to surface processes, which result from the variation of the pulse length of the input power, cannot be well described by the energy density factor. Thus, depending on the introduced power and pulse duration ratio, the same energy density factor results in significantly different properties, e.g., the surface roughness or the chemical layer composition.\cite{kleines2022evaluation}

Again, short pulse durations and low microwave peak power levels have proven advantageous for separating layers in membrane systems, i.e., an energy-deficient process range. For the separation of He from \ce{N$_2$}, the results show a significant decrease of He permeance and He/\ce{N$_2$} selectivity with increasing energy density. The low power prevents strong monomer fragmentation, resulting in less dense and cross-linked growth of the layers. In addition, due to the low pulse durations, rearranging processes are suppressed, and the layer structure does not become denser due to further energy input.\cite{kleines2022evaluation}

With the gained knowledge about the dependence of the PECVD fabrication parameters on the membrane properties, the best performing parameter set of the two mentioned publications of Kleines \textit{et al.} was chosen for the production of membranes for in-depth investigations of the permeation properties of the deposited film \cite{kleines2021structure,kleines2022evaluation}.

Rubner \textit{et al.} \cite{rubner2022mixed} showed that PECVD-fabricated organosilica membranes have the potential to combine the benefits of silica and polymeric membranes, offering both high permeabilities and selectivities comparable to silica membranes and as low production costs as polymeric membranes. The comprehensive investigation of the chosen organosilica membrane, using pure and mixed gas feeds at different temperatures, reveals promising non-ideal permeation behavior. Specifically, the permeance of He and CO$_2$ for the organosilica membrane is strongly temperature-dependent for pure gases, whereas N$_2$ and CH$_4$ permeances remain constant or slightly decrease with increasing temperature. This leads to a significant of size sieving properties with temperature. For mixed gases, Rubner \textit{et al.} studied equimolar mixtures of He/CH$_4$, He/CO$_2$, and CO$_2$/CH$_4$ \cite{rubner2022mixed}. There, the authors observed only minor changes in permeance and mixed gas selectivity compared to the ideal selectivity for the combination of two permanent gases (He and CH$_4$). However, when mixing a condensable gas with He, the membrane exhibits significantly decreased selectivity due to competitive sorption. The same phenomenon also appears to affect the CO$_2$/CH$_4$ selectivity. Rubner \textit{et al.} also observed a significant and irreversible drop in permeance for both He and CO$_2$ when exposed to water vapor, although the ideal He/CO$_2$ selectivity increases \cite{rubner2022mixed}. Overall, the investigated organosilica membrane demonstrates promising mixed gas behavior for separating permanent gases with small kinetic diameters from larger gas molecules.\cite{rubner2022mixed}

Besides the separation properties of the films deposited by PECVD, Rubner \textit{et al.} also investigated the mitigation of physical aging of ultra-thin polymer films with intrinsic microporosity \cite{rubner2022organosilica}. Therefore, they coated the polymer films with a top layer of a few nanometers using PECVD. The study shows that interfacial confinement effects can be used to mitigate aging without altering the intrinsic properties of the polymer. The authors investigated the impact of weakly (SiOCH) and highly (SiO$_x$) oxidized PECVD coating layers on a 200 nm thick polymer with intrinsic microporosity. The results reveal that a SiO$_x$ coating of less than 10 nm can effectively suppress the refractive index increase and hence mitigate physical aging. However, the amplification of the double-sided confinement effect needs to be further investigated to determine the full potential of this approach. The PECVD-coated films exhibit superior anti-aging behavior compared to untreated PTMSP films, which could lead to commercial applications of high free volume polymers with minimized time dependency. Nevertheless, the relationship between aging mitigation and permeation resistance also needs to be studied more closely. Besides, the study also demonstrates an accelerated physical aging of glassy polymers using microwave plasma, which could be used to intentionally age polymers in a fraction of the usual period of time. \\

\subsubsection{Theoretical studies}
\begin{figure}[htb]
\begin{center}
\includegraphics[width=0.4\textwidth]{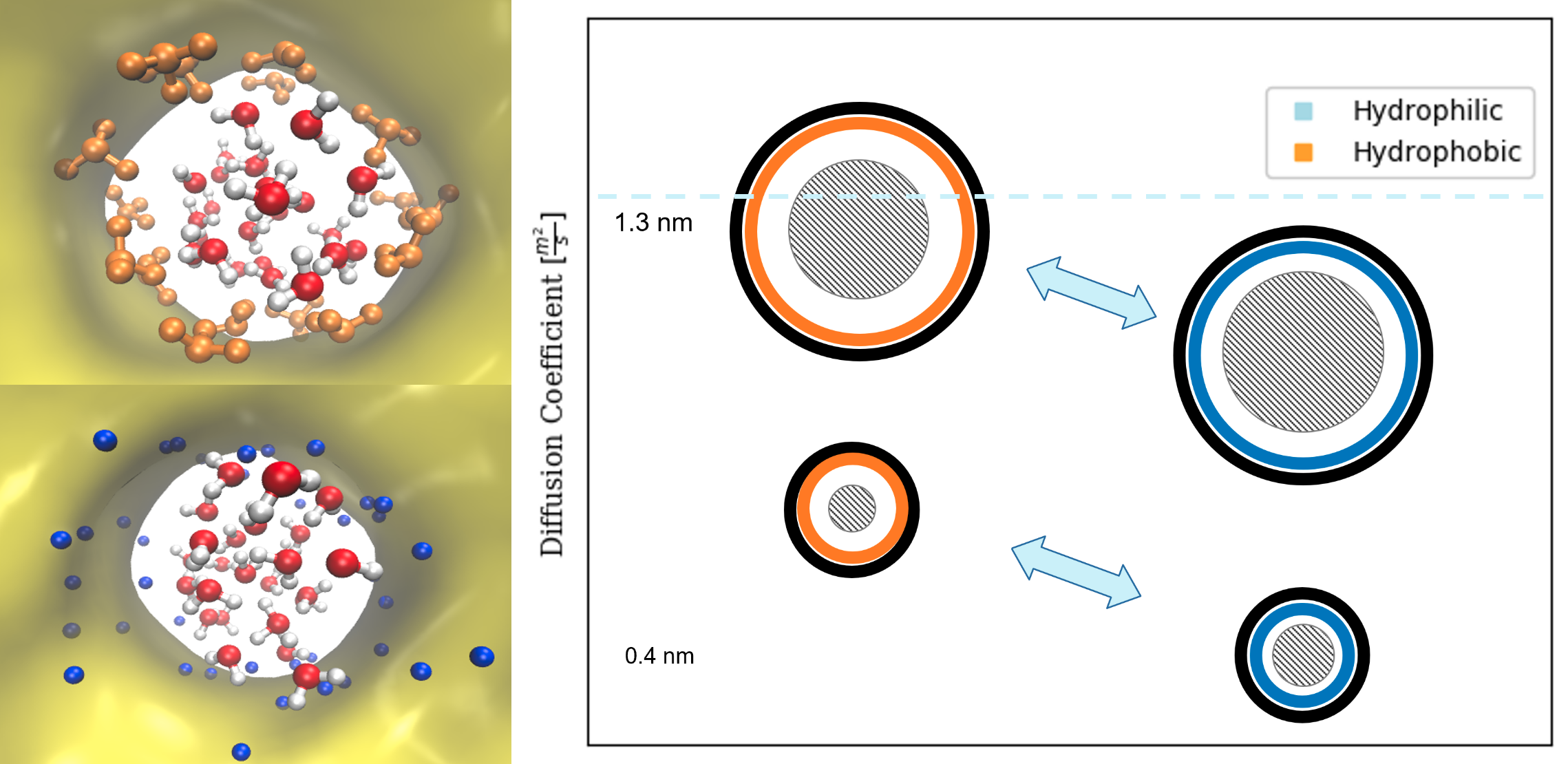}
\end{center}
\caption{Atomistic models of pristine hydrophilic (blue/bottom left) and hydrophobic trifluoromethyl (orange/top left) silica pore systems with water in its interior. The corresponding translational self-diffusion constants are shown in the right panel, which indicates that the diffusion in hydrophobic pores is faster than in equally sized hydrophilic pores and systemically increases with pore size. The diffusion constant of bulk water is denoted by the dashed line for comparison.
\label{fig_DCM_1}}
\end{figure}

The impact of the hydropathy on the transport properties within pores can be best studied by directly comparing idealized hydrophilic and hydrophobic silica pore systems by means of atomistic AIMD simulations. As depicted in Figure~\ref{fig_DCM_1}, pristine hydrophilic systems ranging from 0.35 to 1.37~nm in diameter were investigated, as well as hydrophobic systems with diameters between 0.25 and 1.26~nm that were functionalized with some ratio of trifluoromethyl. Based on our simulations we generally observe a systematically higher translational self-diffusion constant within hydrophobic pores in comparison to their equally sized hydrophilic counterparts, especially for larger pores (see also Figure~\ref{fig_DCM_1}). The fastest diffusion with $2.12\times10^{-9}~m^{2}/s$ was observed in the second largest hydrophilic pore of size 1.27~nm and in the second largest hydrophobic pore of size 1.03~nm with $3.92\times10^{-9}~m^{2}/s$, respectively. Interestingly, the diffusion coefficient in that particular hydrophobic system is not only considerably larger than in all hydrophilic systems we have investigated, but also larger than the experimentally measured diffusion constant in bulk water at ambient conditions ($2.29\times10^{-9}~m^{2}/s$). To verify this and to assess the accuracy of the employed simulation method, the translational self-diffusion of water was calculated with the same semi-empirical method and found to be $2.39\times10^{-9}~m^{2}/s$. Another remarkable observation is that the diffusion of even larger pores becomes smaller again, which indicates that there is only a distinct range of small pore sizes, in which the diffusion is faster than in the bulk. These findings are consistent with the seminar work of Hummer et al., in which they theoretically predicted for the very first time the existence of a burst-like water conduction mechanism within the hydrophobic channel of a carbon nanotube, resulting in water chain moves with little resistance \cite{Hummer2001}.




\subsection{Vapor Phase Infiltration }\label{subsubsec43}

\begin{figure}[htb]
\begin{center}
\includegraphics[width=0.4\textwidth]{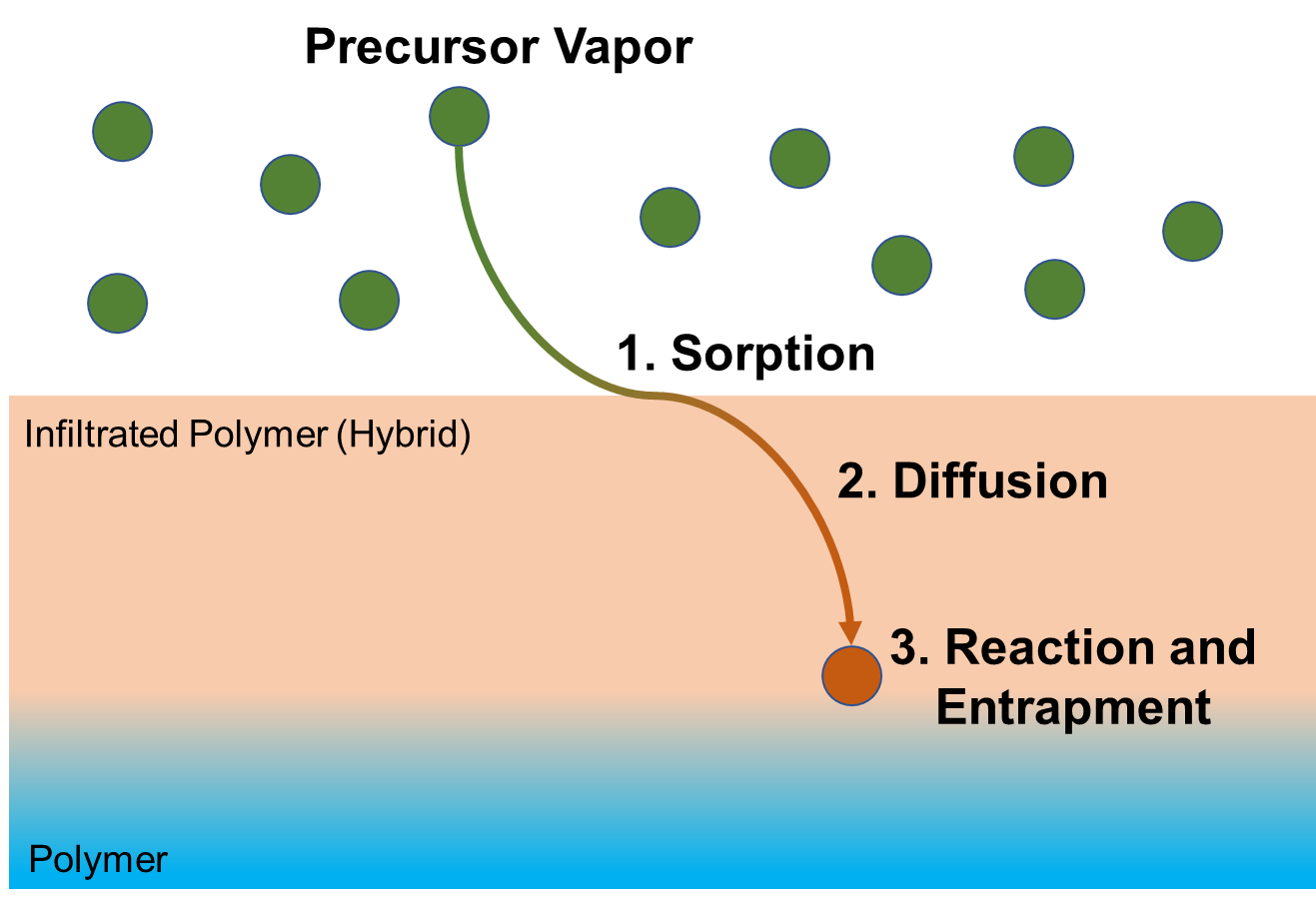}
\end{center}
\caption{Different mechanistic steps influencing the VPI of a precursor vapor into a polymeric material. On the left side is a list with different physico-chemical parameters that are influencing the VPI growth process.  \label{fig_Nils1}}
\end{figure}

The different techniques discussed beforehand, such as PECVD and PEALD, are able to modify the surface by the deposition of a thin film with tailored surface chemistry. This way, pore sizes can be tailored and the inner walls of the pores can be chemically functionalized in order to control the diffusion through the porous material. This is needed in order to control permeation and selectivity of gases through a membrane. However, for very dense and nano-porous materials, PECVD and PEALD are not able to modify the inner walls of the pores completely. This might especially be true for dense polymeric substrates, where the diffusion of precursors and material is proceeding by a solution-diffusion mechanism rather than macroscopic pore diffusion \cite{Wijmans.1995}. To modify the intrinsic pores in these dense polymers and to change the overall functional properties such as permeability and selectivity of gas diffusion, vapor phase infiltration (VPI) can be considered as a valuable method (Figure \ref{fig_Nils1}). \cite{Leng.2017}

\begin{figure*}[htb]
\begin{center}
\includegraphics[width=0.8\textwidth]{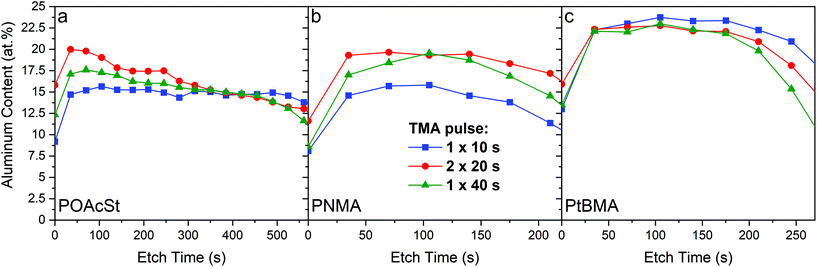}
\end{center}
\caption{Depth profiles obtained from XPS of three different polymers after VPI with different infiltration pulse times. Adapted with permission from the Royal Society of Chemistry \cite{Mai.2022}. \label{fig_Nils2}}
\end{figure*}

Unlike PECVD or PEALD, the gas phase species and precursor molecules are allowed to diffuse inside the polymer by maximizing the exposure of the precursor and the polymer. This can be achieved by using a static rather than a dynamic vacuum while introducing the precursor molecules in the reaction chamber. This way, the precursor molecules have time to diffuse into the confined spaces and dissolve into the polymer. After a defined sorption time, the precursor and other molecular fragments are pumped out of the reactor chamber and a second co-reactant (e.g. water) might be introduced to transform the adsorbed precursor molecules inside the polymer to metal oxides trapped into the polymer. If necessary, this procedure can be repeated in a cyclic fashion in order to control the loading of the polymer with the desired metal oxide and ultimately influence properties such as gas permeability and selectivity. On a mechanistic level at the polymer-vapor interface during VPI processes, a sorption of precursor molecule into the polymer takes place after which the precursor starts to diffuse deeper into the polymer \cite{Leng.2018}. As the precursor molecules and the polymer itself feature reactive functional groups, a reaction of the precursor with the polymer might entrap the precursor in the insides of the polymer. Reaction products and residual precursor molecules can diffuse out of the polymer and are pumped away out of the reaction chamber.

Numerous different applications rely on this special processing technique to form hybrid materials out of organic polymers. Potential applications for the VPI method for hybrid materials are ranging from enhancing the hardness of fibers, over hybrid solar cells, hybrid photoresist materials for lithography in microelectronic fabrication, enhancement of membrane materials and many more \cite{Ashurbekova.2020, Choi.2016, Dwarakanath.2020, Ingram.2019}. Especially for membrane applications, ALD and VPI have proven its effectiveness for the modification of PIM-1 membranes for organic solvent and gas separation \cite{Chen.2021, Liu.2022, McGuinness.2019}. Here, VPI was implemented to specifically tailor the physico-chemical properties of the polymers to transition from gas-barrier polymers to gas-separation membranes. VPI offers another dimension of bulk material conversion compared to PECVD and PEALD, which only offer the functionalization of surfaces in larger pores by thin film deposition.

To gain some preliminary insights, we designed experiments for the infiltration of related but differently functionalized polymers in order to understand the reaction mechanisms that are taking place during infiltration on a molecular level \cite{Mai.2022}. For this, three different polymers, namely \textit{(i)} poly(4-acetoxystyrene) (POAcSt), \textit{(ii)} poly(nonyl methacrylate) (PNMA) and \textit{(iii)} poly(tert-butyl methacrylate) (PtBMA), were spin-coated on silicon substrates and infiltrated afterwards with trimethylaluminum (TMA) which transformed the polymer to Al$_2$O$_3$/organic hybrid films. In this case, water was used as the co-reactant in the VPI process. TMA was introduced into the reaction chamber at low pressure under static vacuum for different pulse durations of 10 s, 2 x 20 s and 40 s, while the spin-coated samples were heated to 85 °C. After the incubation period has passed, residual precursor vapors were pumped away in the vacuum and water was introduced to transform the adsorbed TMA species inside the polymer to Al$_2$O$_3$. The aluminum content of the infiltrated samples at different infiltration times was investigated by XPS. Depth resolution of the Al content was acquired by mild sputtering. Interestingly, the Al-content and distribution after infiltration is different among the different polymer types (Figure \ref{fig_Nils2}).

\begin{figure*}[htb]
\begin{center}
\includegraphics[width=0.8\textwidth]{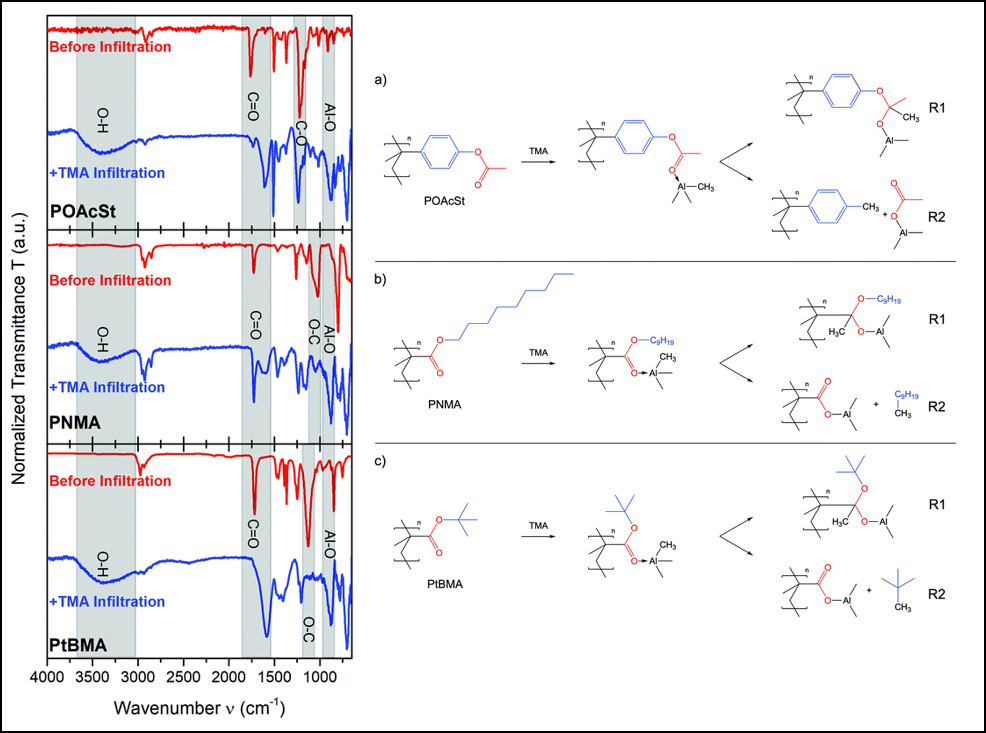}
\end{center}
\caption{\textbf{Left:} FTIR spectra of the different polymers before and after infiltration with TMA for 2 x 20 s at 85 °C. \textbf{Right:} Potential reaction pathways of the polymers after infiltration with TMA. Shown are only the respective monomers. Adapted with permission from the Royal Society of Chemistry \cite{Mai.2022}. \label{fig_Nils3}}
\end{figure*}

For POAcSt and PNMA (Figures \ref{fig_Nils2}a and \ref{fig_Nils2}b), the highest concentration of Al is around 20 at.\% after 2 x 20 s TMA pulse, while for PtBMA (Figure \ref{fig_Nils2}c) around 22.5 at.\% Al could be achieved. Interestingly, the high concentration of 22.5 at.\% Al can be retained through the whole depth of the polymer film in PtBMA, while the distribution of Al is more dependent on the TMA pulse time for POAcSt and especially PNMA. In POAcSt coatings infiltrated by TMA, the Al content seems to be accumulating at the surface of the films and is slightly decreasing to around 15 at.\% after a long sputtering duration.

FTIR spectroscopy was performed with the infiltrated samples in order to gain more information on potential reaction mechanisms that the polymers are undergoing with TMA during infiltration (Figure \ref{fig_Nils3}). Respective spectra were acquired before and after infiltration with 2 x 20 s TMA pulses. For all polymers, a clear contribution of –OH species is visible in the range of 3400 cm$^{-1}$ which indicates the formation of Al–OH species after infiltration and reaction with water. Significant differences are visible in the fingerprint regime for C=O, C–O and Al–O bonds between the different polymer types after infiltration. Especially for POAcSt and PtBMA, a clear shift of the signal at 1750 cm$^{-1}$ towards lower wavenumbers indicates that the C=O bonds in the ester side-chain are converted to bridged centers between two carboxylic groups. The large side chain in PNMA might hinder the diffusion of TMA to the reactive ester in the bulk of the polymer which explains the residual C=O signal at 1750 cm$^{-1}$. This is in line with the results seen in the XPS studies where the Al concentration seems to be highly dependent on the TMA pulsing time which indicates that the overall process might be diffusion limited due to the larger side chains of PNMA.

In all polymers, the signal for the C–O stretches is diminishing, while for PtBMA the signal is completely consumed. This indicates that for PtBMA, a reaction of the ester group with TMA yields a small leaving carbon-based leaving group and nearly all of the C–O bonds are replaced by Al–O bonds. Interestingly, this should explain the high Al content seen in the XPS experiments as the diffusion of TMA seems to be fast and the reaction yields a small leaving group that can easily be pumped out of the polymer and chamber after TMA infiltration.

Based on our experimental findings and experimental results obtained by Dandley \textit{et al.} and Biswas \textit{et al.} \cite{Dandley.2014, Biswas.2014}, different reaction pathways and mechanisms might be assumed which are summarized in Figure \ref{fig_Nils3}. After attractive adsorption of the TMA towards the ester C=O functionality, different reaction pathways are possible. Reaction pathway 1 (R1) yields a hemiacetal by insertion of the methyl group in the C=O bond, while reaction pathway 2 (R2) yields a aluminum functionalized ester group by the elimination of the organic substituent of the original ester for PNMA and PtBMA or by the elimination of a dimehtylaluminumacetate species for POAcSt. Both reaction pathways might be considered as potential outcomes of the infiltration with TMA and can be present simultaneously. As already indicated by the FTIR studies, certain reaction pathways might be more probable to expect than others.

To gain more insights in the reaction pathways and to find the most probable reaction pathway for the infiltration of different polymers with TMA, DFT calculations with the monomeric structures of the different polymers and their interaction and reaction with TMA were carried out. Most interestingly for all polymers, R2 is the most probable reaction pathway to consider from an energetic perspective. If the difference of the energies for the outcome of the reaction $\Delta$E = ER2 - ER1 is considered, it can be estimated to which extent the reactions are following R2 as the most probable reaction pathway. The highest energy difference is observed for PtBMA with –1.497 eV, followed by PNMA with –1.036 eV and POAcSt with –0.456 eV. These results obtained by DFT calculations are in line with the results seen for the XPS measurements and FTIR analysis: The highest Al content that is retained throughout the depth of the polymer film is present in PtBMA. The small organic fragment can easily diffuse out of the polymer and the TMA is strictly bound as a –AlMe2 group to the ester. Also, FTIR indicated a nearly complete conversion of the C–O bonds to Al–O bonds. For POAcSt and PNMA, both reaction pathways might be probable and occurring in the respective infiltration experiments.

As a final conclusion it can be stated that the functional groups and side-chains of the different polymer types are greatly influencing the degree and nature of infiltration with TMA. Thus, it can be assumed that this also drastically influences the applicability of such polymers after infiltration for gas-barrier or membrane applications. In the future, relevant polymers for membrane applications need to be evaluated for their infiltration capabilities in order to precisely tune and optimize their membrane activity in terms of permeability and selectivity. This study only concentrated on the feasibility of VPI into membrane-like polymeric material systems. Besides the VPI of polymers, new studies of our group also indicate that 2D silicate structures fabricated by ALD and strong annealing can be used as membranes for solvent vapor separation \cite{Dementyev.2023, Naberezhnyi.2022}. More detailed studies about the functionalization of potential membrane materials by VPI and ALD are planned in the future.


\section{Experimental and simulation approaches for the transfer and up-scaling of plasma thin film deposition}\label{sec5}
\noindent
Due to the complex nature of plasma processes, development of high quality coatings often takes a long time taking into account many different parameters. Furthermore, the reactors used during this development are not necessarily adapted to industrial needs as the focus is primarily on achieving a successful coating. Transferring a developed coating process thus needs in-depth understanding and can be facilitated by the knowledge of key parameters. In the field of silicon-oxide (SiO$_x$) coatings used as gas barriers for packaging, it has been found that one main factor for achieving a high barrier performance is the existence of a dense, highly cross-linked coating \cite{Deilmann.2008b, Steves.2013}. Using PECVD processes, the cross-linking of the coating can be controlled via the plasma. Usually, such processes are designed by a monomer fed into the reactor, which is polymerised by the plasma forming the obtained film. True to all coating processes is the dependence on the incorporated or consumed energy during deposition as was already indicated in section \ref{subsec41}. In the field of plasma polymerization, this parameter is commonly referred to as energy spent per monomer particle \cite{Hegemann.2021, Starostin.2015, Elam.2017}. More specific towards SiO$_x$ deposition is the introduction of additional oxygen to the process which will then lead to formation of the desired Si-O network. In this case the availability of sufficient amounts of chemically active atomic oxygen is of importance \cite{Ozkaya.2015, Mitschker.2015, Mitschker.2017}.

Mitschker et al. \cite{Mitschker.2018b} investigated the dependence of barrier performance and coating properties of SiO$_x$ coatings from HMDSO-O$_2$ gas mixtures on average ion energy and atomic oxygen flux in relation to the number of incorporated silicon atoms. Using a low pressure MW discharge with attached substrate bias, both parameters could be varied independently allowing for determination of the parameters' specific importance during SiO$_x$ formation.

\begin{figure}[htb]
\begin{center}
\includegraphics[width=0.45\textwidth]{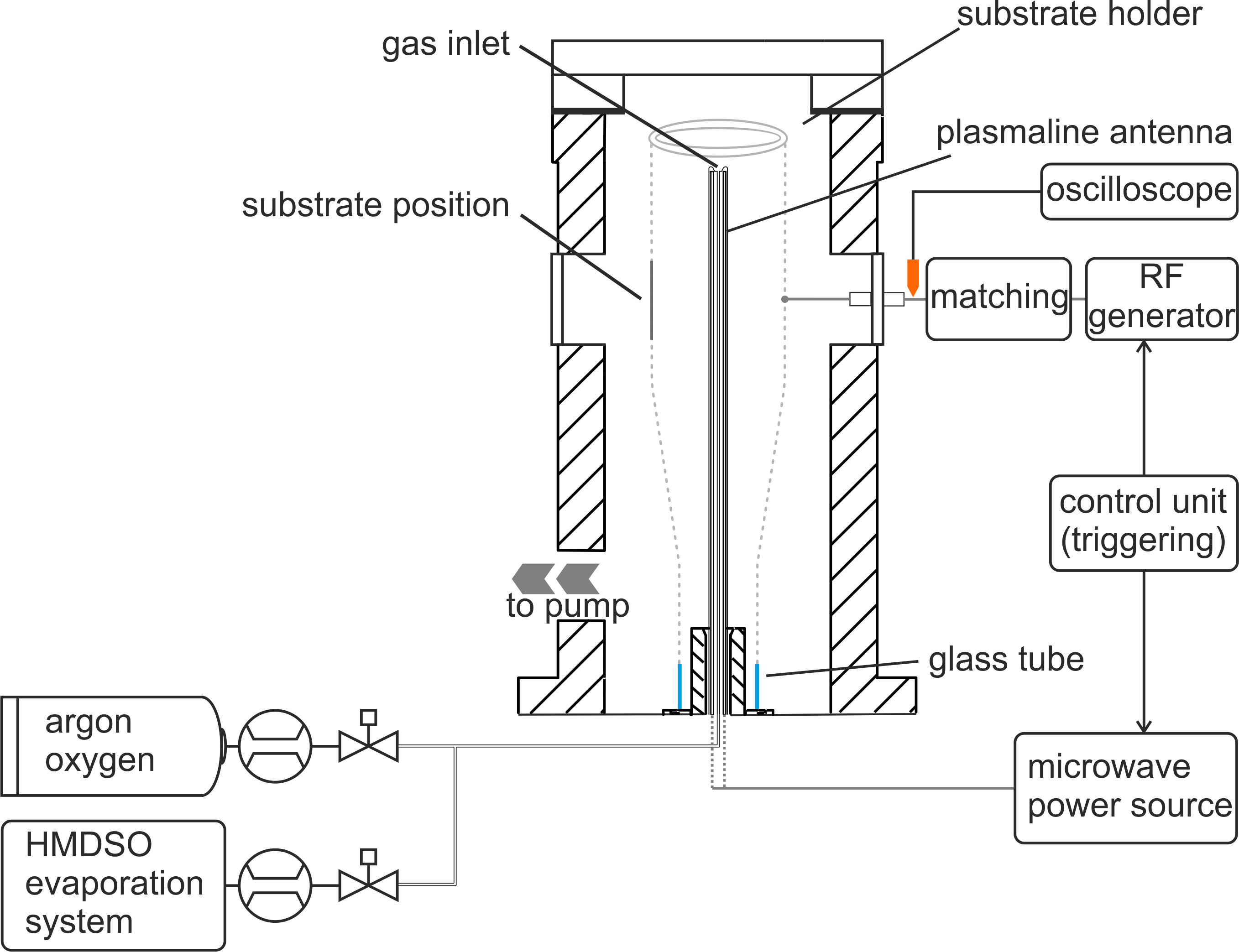}
\end{center}
\caption{Schematic illustration of the experimental setup used for separation of average ion energy and number of atomic oxygen atoms for the deposition of SiO$_x$ coatings. Adapted with permission from IOP Publishing \cite{Mitschker.2018b}.\label{PlasmaLine}}
\end{figure}

A schematic of the reactor used is provided in Figure \ref{PlasmaLine}. It consists of a plasmaline antenna \cite{Petasch.1997, Raeuchle.1998} enclosed by a metallic substrate holder. The plasmaline provides the gas inlet (gas lance) and the microwave power coupling, acting as a coaxial setup when a plasma is ignited. Being originally designed for the coating of PET bottles, the substrate holder, which can either be on ground or equipped with a bias, is designed to hold a 1\,l PET bottle allowing for the inside coating. Aside from this application, the reactor was used for general coatings of polymers and fundamental studies of plasma processes and their relation to coating properties. Due to thermal sensitivity of polymeric substrates, the MW discharge was pulsed.

\begin{figure}[htb]
\begin{center}
\includegraphics[width=0.45\textwidth]{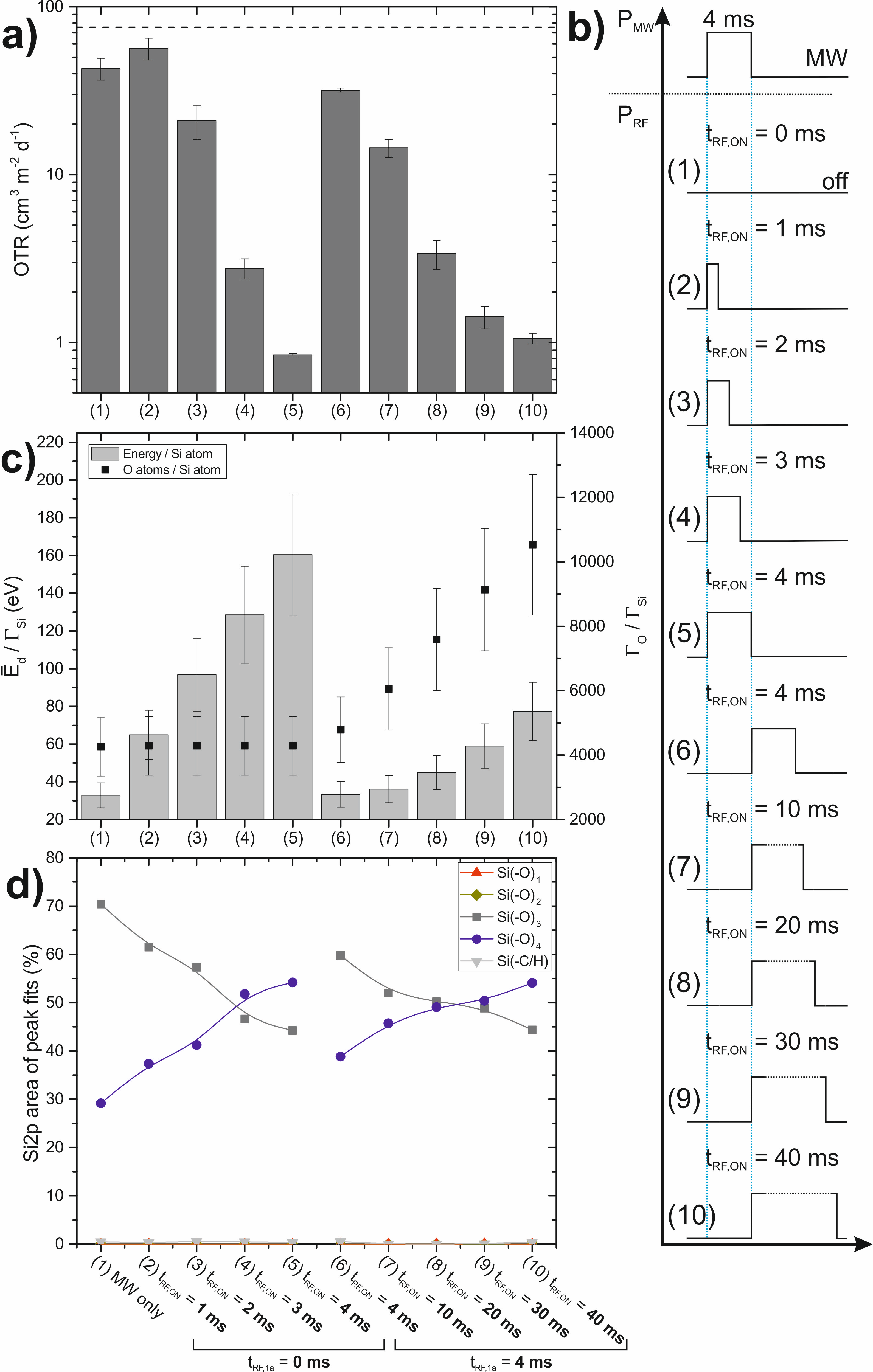}
\end{center}
\caption{\textbf{(a)} Oxygen transmission rate (OTR) of obtained coatings on PET, \textbf{(b)} average incorporated ion energy per Si atom deposited and number of atomic oxygen incident per Si atom deposited, \textbf{(c)} relative area of Si2p components as a function of bias pulse positioning relative to MW pulse, depicted in (b). Reprinted with permission from IOP Publishing \cite{Mitschker.2018b}.}
\label{fig:PulsedBias}
\end{figure}

Positioning of the substrate bias either during MW pulse or MW pause and its corresponding length allowed for separate control of average incorporated ion energy and the incident number of atomic oxygen atoms. Figure \ref{fig:PulsedBias} shows measured average ion energy and number of incident atomic oxygen atoms per deposited Si atom when varying the position of the substrate bias, relative areas of Si2p components (as determined by XPS) in the obtained coatings as well as the resulting OTR for estimation of barrier performance. The pulsed substrate bias was positioned either in the MW discharge (positions (1) to (5)) or during the MW pause (positions (6) to (10)) while varying the length of the bias. Applying a substrate bias during the MW discharge leads to an increase in average ion energy per Si atom deposited from 33\,eV to 160\,eV while keeping the number of atomic oxygen atoms per Si atom deposited at about 4300. When the bias is instead applied during the MW pause, the average ion energy increases only to 77\,eV and the number of atomic oxygen atoms to over 10500. Comparison of relative Si2p areas revealed an increase of Si-O$_4$ components with a decrease of Si-O$_3$ components at the same time, indicating an improved cross-linking for both variations of the substrate bias. With cross-linking being one of the key factors determining the barrier performance, subsequently a reduction in OTR can be seen for the two cases when the coating is deposited on PET substrates.

In the context of transferring SiO$_x$ coatings to industrial scale reactors, it can be concluded in order to achieve a high degree of cross-linking, sufficient fluxes of both ion energy and atomic oxygen with regards to the amount of silicon atoms incorporated needs to be ensured. This is irrespective of plasma source or reactor geometry as is also outlined by Hegemann et al. \cite{Hegemann.2007} for a more generalized macroscopic description of plasma polymerization processes. The same applies to large scale or complex geometries. As long as sufficient fluxes of film forming species, ion energy and reactive oxygen can be ensured at every substrate position, transfer of such coatings is theoretically possible. The process of obtaining in-depth understanding of such a coating process and determination of key parameters can be greatly benefited by the usage of simulations, even more so in the case of limited experimental studies. In terms of transferring and scaling to industrial reactors, simulations are a valuable tool for designing new reactors as fluxes can be estimated beforehand and parts as well as geometry can be adapted accordingly.

We employed two different models: A zero-dimensional ``global model''\cite{kemaneci_2019} to analyze the
complex plasma chemistry,
and a two-dimensional kinetic Particle-in-Cell/Monte Carlo Collisions (PIC/MCC) code\cite{Eremin2022a} to
resolve the underlying plasma   physics  in more detail. Of course, the latter approach
was computationally much more intensive than the former.

\begin{figure}[htb]
\begin{center}
\includegraphics[width=0.45\textwidth]{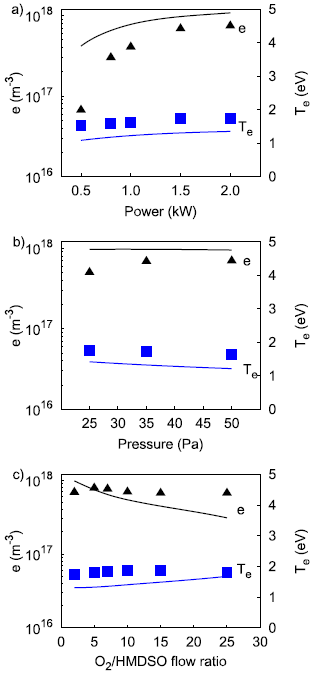}
\end{center}
\caption{Electron density $n_\mathrm{e}$ and electron temperature $T_\mathrm{e}$\linebreak
under variation of a) RF power ($p= 35\,\mathrm{Pa}$,
$Q_{\mathrm{O}_2}/Q_{\mathrm{HMDSO}} = 2$),\linebreak
b) pressure
($P= 1.5\,\mathrm{kW}$, $Q_{\mathrm{O}_2}/Q_{\mathrm{HMDSO}} = 2$),
c) O$_2$/HMDSO flow ratio ($P = 1.5\,\mathrm{kW}$, $p=35\,\mathrm{Pa}$). The simulation
results are denoted by lines, the measurements by points. Reprinted with permission from IOP Publishing \cite{kemaneci_2019}.
}\label{GlobalModelResults}
\end{figure}

The global model employed a self-consistent chemistry set to describe a gas mixture consisting of oxygen
and HMDSO. The model encompassed 34 charged and 69 neutral species and over 1200 chemical reactions, including also surface reactions.\LB
(The reaction rates were collected from a variety of sources.)
The volume-averaged densities of the gas phase species except the electrons was determined by a set of
global particle balance equations which took the co-axial geometry
into account \cite{kemaneci_2017}.
The electron density followed algebraically from the principle of overall charge neutrality.
A Maxwellian electron energy \LB distribution was used to calculate the electron-driven reaction;\LB
the electron temperature was determined by a separate
electron energy balance equation.
The height of the discharge domain, \LB its inner and outer radius, the dissipated power,
the pressure, the composition, and the temperature of the neutral gas were
input parameters guided by experiments.
The study provided a comprehensive analysis of the demonstrator process over a wide parameter range. The simulation results were compared \LB to experimental data
across  various particle density and electron temperature values, yielding agreement concerning the absorbed microwave power, gas pressure, and O$_2$-to-HMDSO flow ratios
(see Fig.~\ref{GlobalModelResults}).

The simulations showed that Si$_2$OC$_5$H$^+_{15}$ dominated the net charge density, with a negligible degree of electronegativity, and played an essential role in the SiO$_x$ deposition process.\LB  We found that HMDSO fragmented into methyl radicals through electron impact dissociation or dissociative ionization, \LB resulting in the generation of large amounts of hydrocarbons such as methane and ethylene,
as well as inorganic molecules like carbon monoxide, carbon dioxide, hydrogen, and water. \LB Significant fractions of the net hydrocarbon and carbon monoxide production were formed by emission from the solid surfaces due to the hydrogen and oxygen atom fluxes.

\begin{figure}[htb]
\begin{center}
\includegraphics[width=0.45\textwidth]{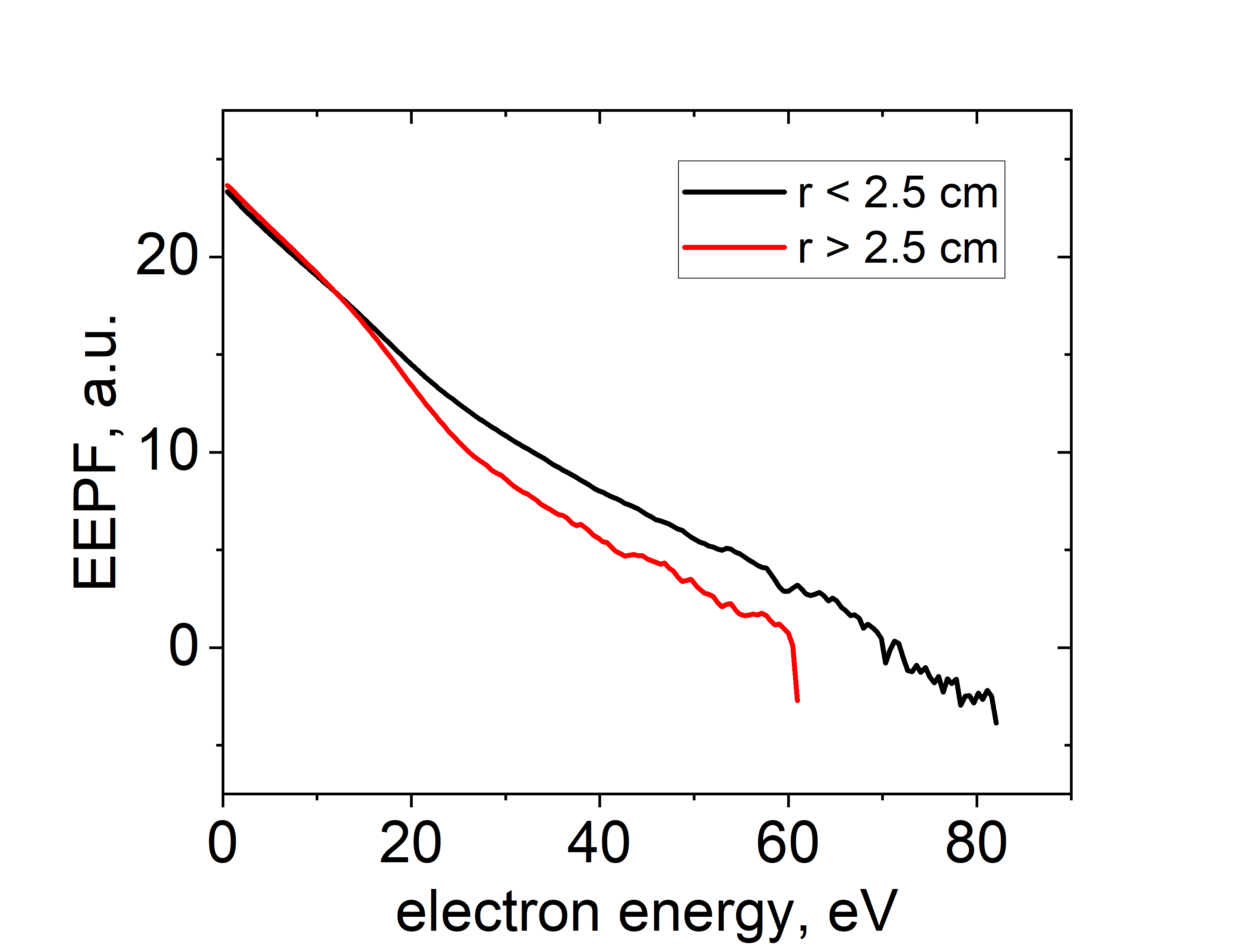}
\end{center}
\caption{Electron energy probability function close to the dielectric surface where
the powering surface wave is excited (black)
and close to the grounded substrate (red). \label{fig_EEPF_plasmaline}}
\end{figure}

The drawback of global models is that they lack temporal and phase space resolution and use many
\textit{a priori} assumptions which makes them not self-consistent. Should this be required, more advanced simulation methods must be used.
One such {\it ab initio} approach is known as Particle-in-Cell/Monte Carlo Collisions (PIC/MCC);
it is particularly suited to study discharges at low pressure where genuine kinetic and nonlocal
effects arise. In the dicharge under study, such effects result from the fact that the
driving frequency of $f= 2.45$ GHz exceeds the collision frequency even at tens of Pa,
and that the energy relaxation length at such pressures, a few centimeters, is comparable to the radial size of the discharge.
This means that the power absorption mechanisms are essentially non-local and must be described kinetically (see Fig.~\ref{fig_EEPF_plasmaline}).


For microwave driving frequencies, an appropriate numerical approach is the electromagnetic particle-in-cell method. Although PIC/MCC as such is already computationally very intensive due to the need to resolve multi-dimensional phase space, its electromagnetic version can be even much more demanding due to limitations stemming from the underlying  algorithms. Such is the case for the simplest explicit scheme, which evolves particles and electromagnetic fields in time using the leapfrog time integration scheme with previously known information. This algorithm becomes unstable if the time needed for a light wave to cross a grid cell is not resolved. This requires very small time steps, not justified by the physics scales of interest. Similarly, the usual explicit momentum-conserving algorithm generates excessive amounts of artificial numerical heating if the Debye length is not resolved by the computation grid, which becomes problematic in dense plasmas such as the one considered. This is why to be able to simulate dense plasmas of large sizes it is necessary to use an implicit algorithm and/or massive plasma parallelization. Both strategies were used to develop in-house family of codes based on the implicit energy- and charge-conserving algorithm for collisional bounded plasmas \cite{Eremin2022} and a 2d generalization of the GPU parallelization approach developed by Mertmann et al.~\cite{mertmann_2011}. The corresponding codes were verified in various benchmarks \cite{turner_2013,charoy_2019,villafana_2021} and were recently validated in studies modeling electrostatic phenomena in rf magnetrons \cite{Berger2022} and electromagnetic phenomena in capacitively coupled plasmas \cite{Eremin2022a}.

\begin{figure}[htb]
\begin{center}
\includegraphics[width=0.4\textwidth]{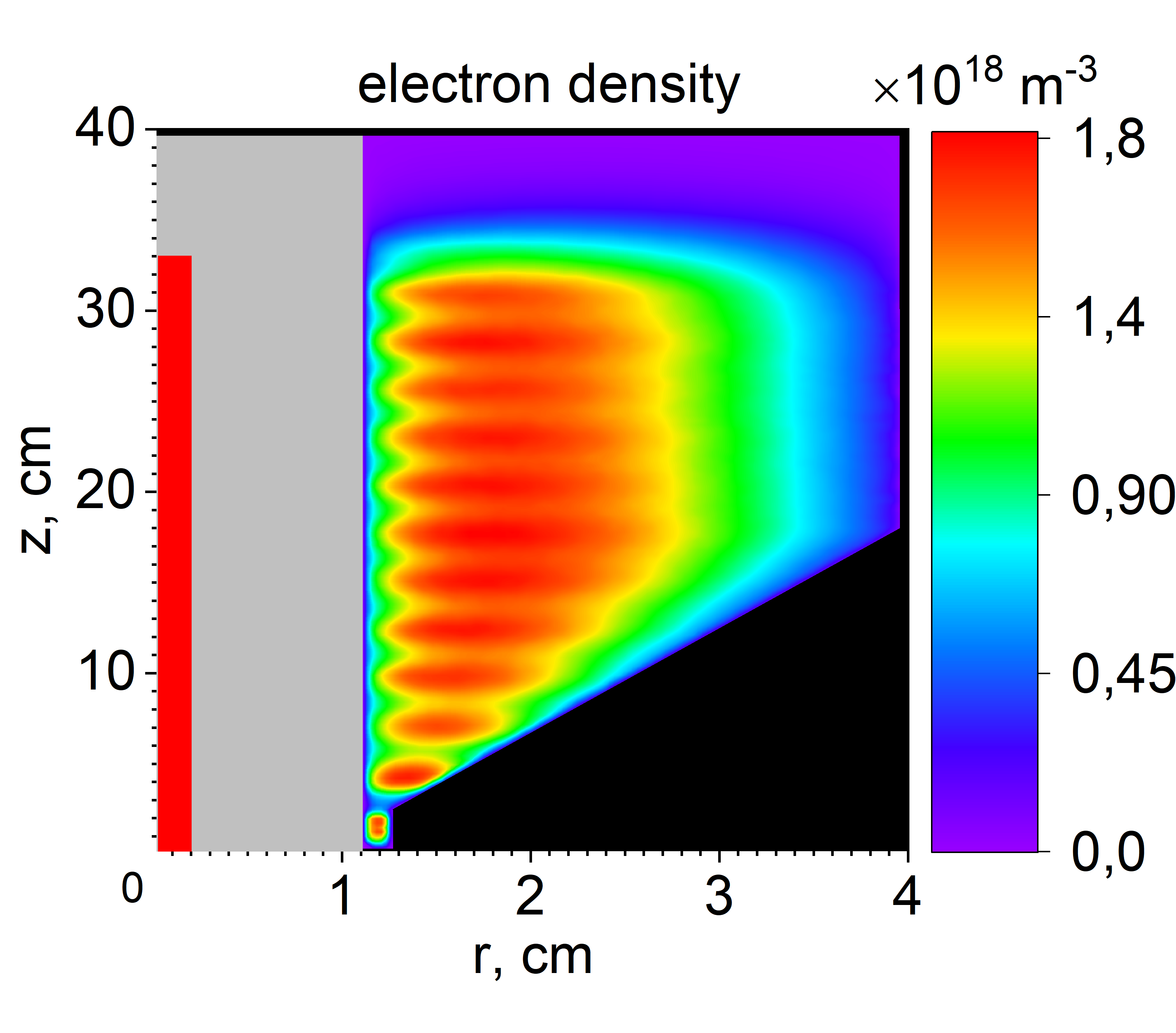}
\end{center}
\caption{Electron densities of the demonstration processs, obtained in a simulation with $10$ Pa and $600$ W. \label{fig_denEl_plasmaline}}
\end{figure}

A cylindrical geometry modeling the plasmaline discharge in question (see Fig.~\ref{PlasmaLine}) is shown in Fig.~\ref{fig_denEl_plasmaline} with a typical electron density profile obtained from a simulation with $10$ Pa and $600$ W. Electromagnetic energy is supplied in the form of a TEM incoming wave between the central rod (red) and the grounded conducting wall (black), which is then converted into a TM surface mode propagating along the boundary between the quartz cylinder (grey) and plasma. The plasma is sustained by the power absorbed from such a surface mode forming a standing wave and exhibits peaks and troughs of the electron density in the axial direction. The electron energy probability function (EEPF) shown in Fig.~\ref{fig_EEPF_plasmaline}, measured in the radial half of the discharge close to the plasmaline, demonstrates a much stronger energetic tail compared to the other discharge half close to the grounded substrate holder, corroborating the need to simulate the discharge kinetically and nonlocally.




\section{Outlook, Perspectives}\label{sec6}


\textbf{Combination of plasma enhanced thin film deposition with complementary surface processes for advanced membrane properties.}  The combination of PEALD and PECVD for enhanced barrier properties as well as the clogging of pores in plasma thin films by PTMSP for improved membrane properties have clearly illustrated that the pores in plasma deposited films can be either sealed e.g. by the PEALD process or a subsequently applied thin polymer film (e.g. PTMSP) actually profits from a given pore structure and surface chemistry.

\textbf{Vapor phase infiltration.}
Vapor phase infiltration of polymers is a promising example for a deposition process of materials inside the free volume of a polymer. The infiltration process changes the chemistry of the polymer from the inside and thus alters the functional properties especially in terms of permeability and selectivity. Especially for gas-barrier and membrane applications, a combination of VPI and PECVD and PEALD processes might be able to further tune the gas-barrier and membrane properties. A combination of both methods allows the polymer to be chemically reinforced from the inside of the polymer and capped on the outside. This does not only change the intrinsic permeability and selectivity of the polymer from the inside but also enables a fine-tuning of the parameters from the outside. The modification by reactive plasmas before or after infiltration might also provide a promising way of further enhancing the selectivity and permeability of the polymers. Especially for unreactive polymers, plasma modification might introduce reactive groups in order to facilitate a reaction of the precursor with the polymer. Despite membrane applications, VPI processing of hybrid materials is gaining significant attention in several other fields. Especially important is the modification of biopolymers, fabrication of hybrid organic-inorganic materials for use in optics, magnetism, photonics, photovoltaics, as diffusion barriers and patterning of materials by photolithography. With a broader understanding of the mechanisms behind the VPI process, this route is expected to extend to niche applications in the future.

Besides VPI, 2D materials such as 2D silicates are gaining a significant interest for membrane applications. In our previous studies we could show that 2D silicates fabricated by ALD, thermal annealing, and transfer can deliver promising membrane activities for vaporized solvents such as methanol, ethanol or isopropanol and gases such as CO$_2$, O$_2$, N$_2$ and noble gases. \cite{naberezhnyi2022, dementyev2022} In the future, this field will be developed further and might become highly interesting for the fabrication of gas-separation membranes with selectivity and permeabilities exceeding the current limitations for classical polymeric membrane systems.

\textbf{New experimental methods for the understanding of pores and gas/pore interactions in thin plasma deposited films.}  Recently, effort has been put into the development of ambient pressure X-ray photoelectron spectroscopy (AP-XPS) for the determination of gas-solid interactions within porous materials. The technical developments that allow to perform XPS in laboratory environments under pressure conditions of up to several kPa have turned the technique into the ideal method for the investigation of gas-solid interactions. As already discussed within this paper, these interactions play a central role for the understanding and control of diffusion processes in thin membranes. Apart from the obvious application of XPS to determine the chemical estate of internal pore walls, the challenge in this case is the detection of effects such as the formation of transient dipoles during physisorption at room temperature between the gas and the pore walls. Through our research it was possible to identify the main challenges associated to such an investigation, which relies in the XPS analysis of the gas phase molecules (as opposed to the analysis of the solid phase). The main characteristic of XPS measurements of the gas phase is the alignment of the electron core levels to the local vacuum levels, so that the measured binding energies of the electrons will shift following any effect that shifts the vacuum levels of the surfaces with which the gas is in contact (such as local electrostatic potentials, work function variations, etc). Therefore, the main issues encountered are two: (i) the electrostatic charging of the samples during the XPS leads to a deformation of the gas peaks; (ii) the gas molecules are spatially distributed in different environments subjected to different local electrostatic potentials (the porous bulk of the sample, the sample surface, and the free gas volume over the sample surface), so that it is difficult to determine what fraction of the detected molecules might be in actual physical contact with the pore walls. The research done until know identifies the issues that need to be taken into account \cite{delosArcos.2022} and has started developing strategies to address these issues \cite{delosArcos.2023}. We believe that the technique will be further developed and that it has the potential to supply information about both chemical and physical interactions between gas and porous materials such as membranes.

Finally, progress must be made in the field of \textbf{modeling and simulation}. As outlined in section 5, we have successfully implemented a volume-averaged plasma chemistry model to simulate the electron density, electron temperature, and the densities of the reactive species.
To achieve a comprehensive description of the demonstrator process, it is imperative to extend the model to incorporate transport processes in a spatially resolved manner. Most importantly,
this applies to the transport of charged particles in the (possibly RF-biased) boundary sheath
over the substrate where energy distributions of the impinging ions are formed.
(Here guidance by the kinetic PIC/MCC simulations will be helpful.)
However, integrating a spatially resolved gas dynamics module may also be essential to account for the behavior of reactive species within the plasma, particular with respect to studying
homogeneity issues. It must be complemented by a chemistry module to capture the complexity of chemical reactions occurring.
In addition, we plan to develop a deposition module that accurately considers both incoming (adsorption) fluxes and outgoing (desorption) fluxes. It is also crucial to describe the formation of the barrier itself, which necessitates the implementation of novel theoretical methods for understanding pores and gas/pore interactions in thin plasma-deposited films. These extensions will be the focus of our future work, which aims to establish a comprehensive understanding of the barrier and membrane formation of thin films on polymer substrates.

\textbf{New theoretical methods for the understanding of pores and gas/pore interactions in thin plasma deposited films.} Second-generation Car-Parrinello-based AIMD simulations unveiled that the self-diffusion in hydrophobic pores is systematically faster than in hydrophilic ones. Moreover, the translational diffusion is increasing with the pore size till it reaches it maximum from where on it decreases again. Nevertheless, there exist a range of pore diameters for which the diffusion constant is even higher than in the bulk. Yet, AIMD simulations on more complex and realistic model systems are needed, which necessitates the use of so called linear-scaling algorithms, whose computational effort is only linearly increasing ith system size \cite{Richters2014,Schade2022}.


\subsection*{Acknowledgments} Funded by the Deutsche Forschungsgemeinschaft (DFG, German Research Foundation) – Project-ID 138690629 – TRR 87.

\subsection*{Data Availability Statement}
The data for this work are available upon reasonable request to the author.

\bibliography{wileyNJD-AMA}

\begin{thebibliography}{100}
\providecommand \doibase [0]{http://dx.doi.org/}%

\bibitem{GiacintiBaschetti.2020}
{Giacinti Baschetti} M, Minelli M. Test methods for the characterization of gas
  and vapor permeability in polymers for food packaging application: A review.
  {\it Polymer Testing} 2020\string; 89\string: 106606.
\newblock \href {\doibase 10.1016/j.polymertesting.2020.106606} {doi:
  10.1016/j.polymertesting.2020.106606}

\bibitem{LeFloch.2018}
{Le Floch} P, Meixuanzi S, Tang J, Liu J, Suo Z. Stretchable Seal. {\it ACS
  Applied Materials {\&} Interfaces} 2018\string; 10(32)\string: 27333--27343.
\newblock \href {\doibase 10.1021/acsami.8b08910} {doi: 10.1021/acsami.8b08910}

\bibitem{McKeen.2017}
McKeen LW. {\it Permeability properties of plastics and elastomers}.
\newblock Oxford: {William Andrew Pub}.
\newblock fourth edition~ed. 2017.

\bibitem{Gaikwad.2018}
Gaikwad KK, Singh S, Lee YS. Oxygen scavenging films in food packaging. {\it
  Environmental Chemistry Letters} 2018\string; 16(2)\string: 523--538.
\newblock \href {\doibase 10.1007/s10311-018-0705-z} {doi:
  10.1007/s10311-018-0705-z}

\bibitem{Bauer.2021}
Bauer AS, Tacker M, Uysal-Unalan I, Cruz RMS, Varzakas T, Krauter V.
  Recyclability and Redesign Challenges in Multilayer Flexible Food Packaging-A
  Review. {\it Foods (Basel, Switzerland)} 2021\string; 10(11).
\newblock \href {\doibase 10.3390/foods10112702} {doi: 10.3390/foods10112702}

\bibitem{Wilski_2020}
Wilski S, Jaritz M, Kleines L, Dahlmann R, Hopmann C. Quantification of
  dominant diffusion processes through plasma enhanced chemical vapor
  deposition-coated plastics by combining two complementary methods for
  porosity analysis. {\it Journal of Physics D: Applied Physics} 2020\string;
  53(32)\string: 325305.
\newblock \href {\doibase 10.1088/1361-6463/ab89cd} {doi:
  10.1088/1361-6463/ab89cd}

\bibitem{sholl2016seven}
Sholl DS, Lively RP. Seven chemical separations to change the world. {\it
  Nature} 2016\string; 532(7600)\string: 435--437.

\bibitem{roualdes20171}
Roualdes S, Rouessac V. 1.10 Plasma Membranes. {\it Comprehensive Membrane
  Science and Engineering} 2017\string: 236--269.

\bibitem{li1999gas}
Li K, Meichsner J. Gas-separating properties of membranes coated by HMDSO
  plasma polymer. {\it Surface and Coatings Technology} 1999\string;
  116\string: 841--847.

\bibitem{lo2010control}
Lo CH, Lin MH, Liao KS, et al. Control of pore structure and characterization
  of plasma-polymerized SiOCH films deposited from octamethylcyclotetrasiloxane
  (OMCTS). {\it Journal of Membrane Science} 2010\string; 365(1-2)\string:
  418--425.

\bibitem{kafrouni2010synthesis}
Kafrouni W, Rouessac V, Julbe A, Durand J. Synthesis and characterization of
  silicon carbonitride films by plasma enhanced chemical vapor deposition
  (PECVD) using bis (dimethylamino) dimethylsilane (BDMADMS), as membrane for a
  small molecule gas separation. {\it Applied Surface Science} 2010\string;
  257(4)\string: 1196--1203.

\bibitem{coustel2014insight}
Coustel R, Haack{\'e} M, Rouessac V, Durand J, Drobek M, Julbe A. An insight
  into the structure--property relationships of PECVD SiCxNy (O): H materials.
  {\it Microporous and mesoporous materials} 2014\string; 191\string: 97--102.

\bibitem{nagasawa2015microporous}
Nagasawa H, Minamizawa T, Kanezashi M, Yoshioka T, Tsuru T. Microporous
  organosilica membranes for gas separation prepared via PECVD using different
  O/Si ratio precursors. {\it Journal of Membrane Science} 2015\string;
  489\string: 11--19.

\bibitem{kleines2020enhancing}
Kleines L, Jaritz M, Wilski S, et al. Enhancing the separation properties of
  plasma polymerized membranes on polydimethylsiloxane substrates by adjusting
  the auxiliary gas in the PECVD processes. {\it Journal of Physics D: Applied
  Physics} 2020\string; 53(44)\string: 445301.

\bibitem{kleines2022evaluation}
Kleines L, Wilski S, Alizadeh P, et al. Evaluation of the membrane performance
  of ultra-smooth silicon organic coatings depending on the process energy
  density. {\it Thin Solid Films} 2022\string; 748\string: 139169.

\bibitem{kafrouni2009synthesis}
Kafrouni W, Rouessac V, Julbe A, Durand J. Synthesis of PECVD a-SiC$_x$N$_y$: H
  membranes as molecular sieves for small gas separation. {\it Journal of
  Membrane Science} 2009\string; 329(1-2)\string: 130--137.

\bibitem{tsuru20112}
Tsuru T, Shigemoto H, Kanezashi M, Yoshioka T. 2-Step plasma-enhanced CVD for
  low-temperature fabrication of silica membranes with high gas-separation
  performance. {\it Chemical Communications} 2011\string; 47(28)\string:
  8070--8072.

\bibitem{nagasawa2013characterization}
Nagasawa H, Shigemoto H, Kanezashi M, Yoshioka T, Tsuru T. Characterization and
  gas permeation properties of amorphous silica membranes prepared via plasma
  enhanced chemical vapor deposition. {\it Journal of membrane science}
  2013\string; 441\string: 45--53.

\bibitem{haacke2015optimization}
Haack{\'e} M, Coustel R, Rouessac V, Drobek M, Rouald{\`e}s S, Julbe A.
  Optimization of the molecular sieving properties of amorphous SiC$_x$N$_y$:H
  hydrogen selective membranes prepared by PECVD. {\it The european physical
  journal special topics} 2015\string; 224(9)\string: 1935--1943.

\bibitem{kleines2021structure}
Kleines L, Wilski S, Alizadeh P, et al. Structure and gas separation properties
  of ultra-smooth PE-CVD silicon organic coated composite membranes. {\it
  Surface and Coatings Technology} 2021\string; 421\string: 127338.

\bibitem{roualdes1999gas}
Roualdes S, Lee V.~dA, Berjoan R, Sanchez J, Durand J. Gas separation
  properties of organosilicon plasma polymerized membranes. {\it AIChE journal}
  1999\string; 45(7)\string: 1566--1575.

\bibitem{roualdes2002gas}
Roualdes S, Sanchez J, Durand J. Gas diffusion and sorption properties of
  polysiloxane membranes prepared by PECVD. {\it Journal of membrane science}
  2002\string; 198(2)\string: 299--310.

\bibitem{bosc2003sorption}
Bosc F, Sanchez J, Rouessac V, Durand J. Sorption and permeation
  characteristics of hybrid organosilicon thin films deposited by PECVD. {\it
  Separation and purification technology} 2003\string; 32(1-3)\string:
  371--376.

\bibitem{ngamou2013plasma}
Ngamou PH, Overbeek JP, Kreiter R, et al. Plasma-deposited hybrid silica
  membranes with a controlled retention of organic bridges. {\it Journal of
  Materials Chemistry A} 2013\string; 1(18)\string: 5567--5576.

\bibitem{charifou2016siox}
Charifou R, Espuche E, Gouanv{\'e} F, Dubost L, Monaco B. SiO$_x$ and
  SiO$_x$C$_z$H$_w$ mono-and multi-layer deposits for improved polymer oxygen
  and water vapor barrier properties. {\it Journal of Membrane Science}
  2016\string; 500\string: 245--254.

\bibitem{wang2017gas}
Wang M, Boscher ND, Heinze K, Gleason KK. Gas selective ultrathin organic
  covalent networks synthesized by iPECVD: does the central metal ion matter?.
  {\it Advanced Functional Materials} 2017\string; 27(29)\string: 1606652.

\bibitem{izu1995roll}
Izu M, Dotter B, Ovshinsky S. Roll-to-roll microwave PECVD machine for
  high-barrier film coatings. {\it Journal of Photopolymer Science and
  Technology} 1995\string; 8(1)\string: 195--204.

\bibitem{baker2012membrane}
Baker RW. {\it Membrane technology and applications}.
\newblock John Wiley \& Sons .
\newblock 2012.

\bibitem{rossi1993}
Rossi G, Nulman M. Effect of local flaws in polymeric permeation reducing
  barriers. {\it Journal of Applied Physics} 1993\string; 74(9)\string:
  5471-5475.
\newblock \href {\doibase 10.1063/1.354227} {doi: 10.1063/1.354227}

\bibitem{dasilva2000}
Silva~Sobrinho dAS, Czeremuszkin G, Latrèche M, Wertheimer MR.
  Defect-permeation correlation for ultrathin transparent barrier coatings on
  polymers. {\it Journal of Vacuum Science \& Technology A} 2000\string;
  18(1)\string: 149-157.
\newblock \href {\doibase 10.1116/1.582156} {doi: 10.1116/1.582156}

\bibitem{hanika2003}
Hanika M, Langowski HC, Moosheimer U, Peukert W. Inorganic Layers on Polymeric
  Films – Influence of Defects and Morphology on Barrier Properties. {\it
  Chemical Engineering \& Technology} 2003\string; 26(5)\string: 605-614.
\newblock \href {\doibase doi.org/10.1002/ceat.200390093} {doi:
  doi.org/10.1002/ceat.200390093}

\bibitem{Wilski_2017}
Wilski S, Wipperfürth J, Jaritz M, et al. Mechanisms of oxygen permeation
  through plastic films and barrier coatings. {\it Journal of Physics D:
  Applied Physics} 2017\string; 50(42)\string: 425301.
\newblock \href {\doibase 10.1088/1361-6463/aa8525} {doi:
  10.1088/1361-6463/aa8525}

\bibitem{Wilski_2021}
Wilski S, Alizadeh P, Kleines L, Hopmann C, Dahlmann R. Influence of pore
  spacing in barrier coatings on the mass transport through plastics—a
  simulative and experimental approach. {\it Journal of Physics D: Applied
  Physics} 2021\string; 54(47)\string: 475305.
\newblock \href {\doibase 10.1088/1361-6463/ac2385} {doi:
  10.1088/1361-6463/ac2385}

\bibitem{Kuehne2014}
Kühne TD. Second generation Car-Parrinello molecular dynamics. {\it WIREs
  Comput. Mol. Sci.} 2014\string; 4(4)\string: 391-404.
\newblock \href {\doibase 10.1002/wcms.1176} {doi: 10.1002/wcms.1176}

\bibitem{Kuehne2015}
Spura T, John C, Haberhson S, Kühne TD. Nuclear quantum effects in liquid
  water from path-integral simulations using an ab initio force-matching
  approach. {\it Mol. Phys.} 2015\string; 113(8)\string: 802-822.
\newblock \href {\doibase 10.1080/00268976.2014.981231} {doi:
  10.1080/00268976.2014.981231}

\bibitem{Ghasemi2022}
Khajehpasha ER, Finkler JA, Kühne TD, Ghasemi SA. Nuclear quantum effects in
  liquid water from path-integral simulations using an ab initio force-matching
  approach. {\it Phys. Rev. B} 2022\string; 105(14)\string: 144106.
\newblock \href {\doibase 10.1103/PhysRevB.105.144106} {doi:
  10.1103/PhysRevB.105.144106}

\bibitem{Hutter2023}
Hutter J, Iannuzzi M, Kühne TD. Ab Initio Molecular Dynamics: A Guide to
  Applications. In: Elsevier.  2023

\bibitem{Car1985}
Car R, Parrinello M. Unified Approach for Molecular Dynamics and
  Density-Functional Theory. {\it Phys. Rev. Lett.} 1985\string; 55(22)\string:
  2471.
\newblock \href {\doibase 10.1103/PhysRevLett.55.2471} {doi:
  10.1103/PhysRevLett.55.2471}

\bibitem{Kuehne2007}
Kühne TD, Krack M, Mohammad FR, Parrinello M. Efficient and Accurate
  Car-Parrinello-like Approach to Born-Oppenheimer Molecular Dynamics. {\it
  Phys. Rev. Lett.} 2007\string; 98(6)\string: 066401.
\newblock \href {\doibase 10.1103/PhysRevLett.98.066401} {doi:
  10.1103/PhysRevLett.98.066401}

\bibitem{Prodan2018}
Kühne TD, Prodan E. Disordered crystals from first principles I: Quantifying
  the configuration space. {\it Annals of Physics} 2018\string; 391\string:
  120-149.
\newblock \href {\doibase 10.1016/j.aop.2018.01.016} {doi:
  10.1016/j.aop.2018.01.016}

\bibitem{Prodan2020}
Kühne TD, Heske J, Prodan E. Disordered crystals from first principles II:
  Transport coefficients. {\it Annals of Physics} 2020\string; 421\string:
  168290.
\newblock \href {\doibase 10.1016/j.aop.2020.168290} {doi:
  10.1016/j.aop.2020.168290}

\bibitem{Camellone2009}
Camellone MF, Kühne TD, Passerone D. Density functional theory study of
  self-trapped holes in disordered SiO2. {\it Phys. Rev. B} 2009\string;
  80(3)\string: 033203.
\newblock \href {\doibase 10.1103/PhysRevB.80.033203} {doi:
  10.1103/PhysRevB.80.033203}

\bibitem{Allolio_2014}
Allolio C, Klameth F, Vogel M, Sebastiani D. Ab initio H$_2$O in realistic
  hydrophilic confinement. {\it ChemPhysChem} 2014\string; 15(18)\string:
  3955--3962.

\bibitem{Gallo2010}
Gallo P, Rovere M, Chen SH. Anomalous dynamics of water confined in MCM-41 at
  different hydrations. {\it Journal of Physics: Condensed Matter} 2010\string;
  22(28)\string: 284102.
\newblock \href {\doibase 10.1088/0953-8984/22/28/284102} {doi:
  10.1088/0953-8984/22/28/284102}

\bibitem{Weinberger2016}
Weinberger C, Vetter S, Tiemann M, Wagner T. Assessment of the density of
  (meso)porous materials from standard volumetric physisorption data. {\it
  Microporous and Mesoporous Materials} 2016\string; 223\string: 53-57.
\newblock \href {\doibase doi.org/10.1016/j.micromeso.2015.10.027} {doi:
  doi.org/10.1016/j.micromeso.2015.10.027}

\bibitem{hartnig2000}
Hartnig C, Witschel W, Spohr E, Gallo P, Ricci MA, Rovere M. Modifications of
  the hydrogen bond network of liquid water in a cylindrical SiO$_2$ pore. {\it
  Journal of Molecular Liquids} 2000\string; 85(1-2)\string: 127--137.

\bibitem{Nakano1994}
Nakano A, Bi L, Kalia RK, Vashishta P. Molecular-dynamics study of the
  structural correlation of porous silica with use of a parallel computer. {\it
  Phys. Rev. B} 1994\string; 49\string: 9441--9452.
\newblock \href {\doibase 10.1103/PhysRevB.49.9441} {doi:
  10.1103/PhysRevB.49.9441}

\bibitem{Kieffer1988}
Kieffer J, Angell C. Generation of fractal structures by negative pressure
  rupturing of SiO$_2$ glass. {\it Journal of Non-Crystalline Solids}
  1988\string; 106(1)\string: 336-342.
\newblock \href {\doibase doi.org/10.1016/0022-3093(88)90286-4} {doi:
  doi.org/10.1016/0022-3093(88)90286-4}

\bibitem{Gergs2022}
Gergs T, Monti C, Gaiser S, et al. Nanoporous SiO$_x$ plasma polymer films as
  carrier for liquid-infused surfaces. {\it Plasma Processes and Polymers}
  2022\string; 19(8)\string: 2200049.
\newblock \href {\doibase https://doi.org/10.1002/ppap.202200049} {doi:
  https://doi.org/10.1002/ppap.202200049}

\bibitem{mosaddeghi2012}
Mosaddeghi H, Alavi S, Kowsari M, Najafi B. Simulations of structural and
  dynamic anisotropy in nano-confined water between parallel graphite plates.
  {\it The Journal of chemical physics} 2012\string; 137(18)\string: 184703.

\bibitem{ghoufi2011}
Ghoufi A, Morineau D, Lefort R, et al. Molecular simulations of confined
  liquids: An alternative to the grand canonical Monte Carlo simulations. {\it
  The Journal of chemical physics} 2011\string; 134(7)\string: 074104.

\bibitem{shroll1999molecular}
Shroll RM, Smith DE. Molecular dynamics simulations in the grand canonical
  ensemble: Application to clay mineral swelling. {\it The Journal of chemical
  physics} 1999\string; 111(19)\string: 9025--9033.

\bibitem{shevade2000molecular}
Shevade AV, Jiang S, Gubbins KE. Molecular simulation study of water--methanol
  mixtures in activated carbon pores. {\it The Journal of Chemical Physics}
  2000\string; 113(16)\string: 6933--6942.

\bibitem{Lerbret_2011}
Lerbret A, Lelong G, Mason PE, Saboungi ML, Brady JW. Water confined in
  cylindrical pores: a molecular dynamics study. {\it Food biophysics}
  2011\string; 6(2)\string: 233--240.

\bibitem{Bourg_2012}
Bourg IC, Steefel CI. Molecular dynamics simulations of water structure and
  diffusion in silica nanopores. {\it The Journal of Physical Chemistry C}
  2012\string; 116(21)\string: 11556--11564.

\bibitem{Renou_2014}
Renou R, Szymczyk A, Ghoufi A. Water confinement in nanoporous silica
  materials. {\it The Journal of Chemical Physics} 2014\string; 140(4)\string:
  044704.

\bibitem{Gallo_2012}
Gallo P, Rovere M, Chen S. Water confined in MCM-41: a mode coupling theory
  analysis. {\it Journal of Physics: Condensed Matter} 2012\string;
  24(6)\string: 064109.

\bibitem{Sulpizi_2012}
Sulpizi M, Gaigeot MP, Sprik M. The silica--water interface: how the silanols
  determine the surface acidity and modulate the water properties. {\it Journal
  of chemical theory and computation} 2012\string; 8(3)\string: 1037--1047.

\bibitem{Cimas_2014}
Cimas {\'{A}}, Tielens F, Sulpizi M, Gaigeot MP, Costa D. The amorphous
  silica{\textendash}liquid water interface studied by ab initio molecular
  dynamics ({AIMD}): local organization in global disorder. {\it Journal of
  Physics: Condensed Matter} 2014\string; 26(24)\string: 244106.
\newblock \href {\doibase 10.1088/0953-8984/26/24/244106} {doi:
  10.1088/0953-8984/26/24/244106}

\bibitem{Weinberger_2022}
Weinberger C, Zysk F, Hartmann M, et al. The Structure of Water in Silica
  Mesopores – Influence of the Pore Wall Polarity. {\it Advanced Materials
  Interfaces} 2022\string; 9(20)\string: 2200245.
\newblock \href {\doibase doi.org/10.1002/admi.202200245} {doi:
  doi.org/10.1002/admi.202200245}

\bibitem{Voorhis2015}
Welborn M, Chen J, Wang LP, Voorhis TV. Why many semiempirical molecular
  orbital theories fail for liquid water and how to fix them. {\it J. Comp.
  Chem.} 2015\string; 36(12)\string: 934-939.
\newblock \href {\doibase 10.1002/jcc.23887} {doi: 10.1002/jcc.23887}

\bibitem{Kuo.2015}
Kuo YL, Chang KH. Atmospheric pressure plasma enhanced chemical vapor
  deposition of SiO$_x$ films for improved corrosion resistant properties of
  AZ31 magnesium alloys. {\it Surface and Coatings Technology} 2015\string;
  283\string: 194--200.
\newblock \href {\doibase 10.1016/j.surfcoat.2015.11.004} {doi:
  10.1016/j.surfcoat.2015.11.004}

\bibitem{Alissawi.2013}
Alissawi N, Peter T, Strunskus T, Ebbert C, Grundmeier G, Faupel F.
  Plasma-polymerized HMDSO coatings to adjust the silver ion release properties
  of Ag/polymer nanocomposites. {\it Journal of Nanoparticle Research}
  2013\string; 15(11)\string: 1--12.
\newblock \href {\doibase 10.1007/s11051-013-2080-9} {doi:
  10.1007/s11051-013-2080-9}

\bibitem{Perrotta.2015}
Perrotta A, Garc{\'i}a SJ, Creatore M. Ellipsometric Porosimetry and
  Electrochemical Impedance Spectroscopy Characterization for Moisture
  Permeation Barrier Layers. {\it Plasma Processes and Polymers} 2015\string;
  12(9)\string: 968--979.
\newblock \href {\doibase 10.1002/ppap.201500084} {doi: 10.1002/ppap.201500084}

\bibitem{Perrotta.2015b}
Perrotta A, Garc{\'i}a SJ, Michels JJ, Andringa AM, Creatore M. Analysis of
  Nanoporosity in Moisture Permeation Barrier Layers by Electrochemical
  Impedance Spectroscopy. {\it ACS applied materials {\&} interfaces}
  2015\string; 7(29)\string: 15968--15977.
\newblock \href {\doibase 10.1021/acsami.5b04060} {doi: 10.1021/acsami.5b04060}

\bibitem{Perrotta.2016}
Perrotta A, Kessels WMM, Creatore M. Dynamic Ellipsometric Porosimetry
  Investigation of Permeation Pathways in Moisture Barrier Layers on Polymers.
  {\it ACS applied materials {\&} interfaces} 2016\string; 8(38)\string:
  25005--25009.
\newblock \href {\doibase 10.1021/acsami.6b08520} {doi: 10.1021/acsami.6b08520}

\bibitem{Xie.2022}
Xie X, {de los Arcos} T, Grundmeier G. {Comparative analysis of
  hexamethyldisiloxane and hexamethyldisilazane plasma polymer thin films
  before and after plasma oxidation}. {\it {Plasma Processes and Polymers}}
  2022.
\newblock \href {\doibase 10.1002/ppap.202200052} {doi: 10.1002/ppap.202200052}

\bibitem{Liu_2003}
Liu YC, Wang Q, Lu LH. Water confined in nanopores: its molecular distribution
  and diffusion at lower density. {\it Chemical physics letters} 2003\string;
  381(1-2)\string: 210--215.

\bibitem{Ohto_2015}
Ohto T, Usui K, Hasegawa T, Bonn M, Nagata Y. Toward ab initio molecular
  dynamics modeling for sum-frequency generation spectra; an efficient
  algorithm based on surface-specific velocity-velocity correlation function.
  {\it The Journal of Chemical Physics} 2015\string; 143(12)\string: 124702.

\bibitem{Hoppe.2017}
Hoppe C, Mitschker F, Giner I, {de los Arcos} T, Awakowicz P, Grundmeier G.
  Influence of organic surface chemistry on the nucleation of plasma deposited
  SiO$_x$ films. {\it Journal of Physics D: Applied Physics} 2017\string;
  50(20)\string: 204002.
\newblock \href {\doibase 10.1088/1361-6463/aa69e5} {doi:
  10.1088/1361-6463/aa69e5}

\bibitem{Hoppe.2020}
Hoppe C, Mitschker F, Butterling M, et al. {Characterisation of micropores in
  plasma deposited SiO$_x$ films by means of positron annihilation lifetime
  spectroscopy}. {\it Journal of Physics D: Applied Physics} 2020\string;
  53(47)\string: 475205.
\newblock \href {\doibase 10.1088/1361-6463/aba8ba} {doi:
  10.1088/1361-6463/aba8ba}

\bibitem{Hoppe.2022}
Hoppe C, Mitschker F, Mai L, et al. {Influence of surface activation on the
  microporosity of PE--CVD and PE--ALD SiO$_x$ thin films on PDMS}. {\it
  {Plasma Processes and Polymers}} 2022.
\newblock \href {\doibase 10.1002/ppap.202100174} {doi: 10.1002/ppap.202100174}

\bibitem{delosArcos.2021}
{de los Arcos} T, M{\"u}ller H, Wang F, et al. {Review of infrared spectroscopy
  techniques for the determination of internal structure in thin SiO$_2$
  films}. {\it {Vibrational Spectroscopy}} 2021\string; 114\string: 103256.
\newblock \href {\doibase 10.1016/j.vibspec.2021.103256} {doi:
  10.1016/j.vibspec.2021.103256}

\bibitem{Menzel.2001}
Menzel O. {\it Charakterisierung makropor{\"o}ser Materialien mit den Methoden
  der digitalen Bildverarbeitung und Bestimmung von effektiven
  Diffusionskoeffizienten durch Computersimulationen}. PhD thesis. {Hannover :
  Gottfried Wilhelm Leibniz Universit{\"a}t Hannover},  2001

\bibitem{Wilski.2022}
Wilski S. {\it Analyse der Porenstrukturen in nanostrukturierten
  Funktionsschichten auf Kunststoffen und Modellierung des porengesteuerten
  Stofftransports}. Dissertation. {Verlag Mainz} and
  {Rheinisch-Westf{\"a}lische Technische Hochschule Aachen},  2022.

\bibitem{Bahroun_2014}
Bahroun K, Behm H, Mitschker F, Awakowicz P, Dahlmann R, Hopmann C. Influence
  of layer type and order on barrier properties of multilayer PECVD barrier
  coatings. {\it Journal of Physics D: Applied Physics} 2013\string;
  47(1)\string: 015201.
\newblock \href {\doibase 10.1088/0022-3727/47/1/015201} {doi:
  10.1088/0022-3727/47/1/015201}

\bibitem{Mitschker.2018}
Mitschker F, Schuecke L, Hoppe C, et al. {Comparative study on the deposition
  of silicon oxide permeation barrier coatings for polymers using
  hexamethyldisilazane (HMDSN) and hexamethyldisiloxane (HMDSO)}. {\it Journal
  of Physics D: Apllied Physics} 2018\string; 51(23).
\newblock \href {\doibase 10.1088/1361-6463/aac0ab} {doi:
  10.1088/1361-6463/aac0ab}

\bibitem{Jaritz.2017}
Jaritz M, Hopmann C, Behm H, et al. Influence of Residual Stress on the
  Adhesion and Surface Morphology of {{PECVD-coated}} Polypropylene. {\it
  Journal of Physics D: Applied Physics} 2017\string; 50(44)\string: 445301.
\newblock \href {\doibase 10.1088/1361-6463/ aa8798} {doi: 10.1088/1361-6463/
  aa8798}

\bibitem{jaritz2021}
Jaritz M, Alizadeh P, Wilski S, Kleines L, Dahlmann R. Comparison of HMDSO and
  HMDSN as precursors for high-barrier plasma-polymerized multilayer coating
  systems on polyethylene terephthalate films. {\it Plasma Processes and
  Polymers} 2021\string; 18(8)\string: 2100018.
\newblock \href {\doibase doi.org/10.1002/ppap.202100018} {doi:
  doi.org/10.1002/ppap.202100018}

\bibitem{Behm.2014}
Behm H, Bahroun K, Bahre H, et al. Adhesion of {{Thin CVD Films}} on {{Pulsed
  Plasma Pre-Treated Polypropylene}}: {{Adhesion}} of {{Thin CVD Films}} on
  {{Pulsed Plasma}} \ldots. {\it Plasma Processes and Polymers} 2014\string;
  11(5)\string: 418--425.
\newblock \href {\doibase 10.1002/ppap.201300128} {doi: 10.1002/ppap.201300128}

\bibitem{Jaritz.2017b}
Jaritz M, Behm H, Hopmann C, et al. The effect of UV radiation from oxygen and
  argon plasma on the adhesion of organosilicon coatings on polypropylene. {\it
  Journal of Physics D: Applied Physics} 2016\string; 50(1)\string: 015201.
\newblock \href {\doibase 10.1088/1361-6463/50/1/015201} {doi:
  10.1088/1361-6463/50/1/015201}

\bibitem{yave2010nanometric}
Yave W, Car A, Wind J, Peinemann KV. Nanometric thin film membranes
  manufactured on square meter scale: ultra-thin films for CO$_x$ capture. {\it
  Nanotechnology} 2010\string; 21(39)\string: 395301.

\bibitem{jaritz2020hmdso}
Jaritz M, Hopmann C, Wilski S, et al. HMDSO-based thin plasma polymers as
  corrosion barrier against NaOH solution. {\it Journal of Materials
  Engineering and Performance} 2020\string; 29(5)\string: 2839--2847.

\bibitem{rubner2022mixed}
Rubner J, Skribbe S, Roth H, Kleines L, Dahlmann R, Wessling M. On the Mixed
  Gas Behavior of Organosilica Membranes Fabricated by Plasma-Enhanced Chemical
  Vapor Deposition (PECVD). {\it Membranes} 2022\string; 12(10)\string: 994.

\bibitem{rubner2022organosilica}
Rubner J, Stellmann L, Mertens AK, et al. Organosilica coating layer prevents
  aging of a polymer with intrinsic microporosity. {\it Plasma Processes and
  Polymers} 2022\string; 19(8)\string: 2200016.

\bibitem{Hummer2001}
Hummer G, Rasaiah JC, Noworyta JP. Water conduction through the hydrophobic
  channel of a carbon nanotube. {\it Nature} 2001\string; 414\string:
  188–190.
\newblock \href {\doibase 10.1038/35102535} {doi: 10.1038/35102535}

\bibitem{Wijmans.1995}
Wijmans JG, Baker RW. The solution-diffusion model: a review. {\it Journal of
  Membrane Science} 1995\string; 107(1-2)\string: 1--21.
\newblock \href {\doibase 10.1016/0376-7388(95)00102-I} {doi:
  10.1016/0376-7388(95)00102-I}

\bibitem{Leng.2017}
Leng CZ, Losego MD. Vapor phase infiltration (VPI) for transforming polymers
  into organic--inorganic hybrid materials: a critical review of current
  progress and future challenges. {\it Materials Horizons} 2017\string;
  4(5)\string: 747--771.
\newblock \href {\doibase 10.1039/C7MH00196G} {doi: 10.1039/C7MH00196G}

\bibitem{Mai.2022}
Mai L, Maniar D, Zysk F, et al. Influence of different ester side groups in
  polymers on the vapor phase infiltration with trimethyl aluminum. {\it Dalton
  transactions (Cambridge, England : 2003)} 2022\string; 51(4)\string:
  1384--1394.
\newblock \href {\doibase 10.1039/D1DT03753F} {doi: 10.1039/D1DT03753F}

\bibitem{Leng.2018}
Leng CZ, Losego MD. A physiochemical processing kinetics model for the vapor
  phase infiltration of polymers: measuring the energetics of precursor-polymer
  sorption, diffusion, and reaction. {\it Physical Chemistry Chemical Physics}
  2018\string; 20(33)\string: 21506--21514.
\newblock \href {\doibase 10.1039/C8CP04135K} {doi: 10.1039/C8CP04135K}

\bibitem{Ashurbekova.2020}
Ashurbekova K, Ashurbekova K, Botta G, Yurkevich O, Knez M. Vapor phase
  processing: a novel approach for fabricating functional hybrid materials.
  {\it Nanotechnology} 2020\string; 31(34)\string: 342001.
\newblock \href {\doibase 10.1088/1361-6528/ab8edb} {doi:
  10.1088/1361-6528/ab8edb}

\bibitem{Choi.2016}
Choi JW, Li Z, Black CT, Sweat DP, Wang X, Gopalan P. Patterning at the 10
  nanometer length scale using a strongly segregating block copolymer thin film
  and vapor phase infiltration of inorganic precursors. {\it Nanoscale}
  2016\string; 8(22)\string: 11595--11601.
\newblock \href {\doibase 10.1039/C6NR01409G} {doi: 10.1039/C6NR01409G}

\bibitem{Dwarakanath.2020}
Dwarakanath S, Raj PM, Kondekar N, Losego MD, Tummala R. Vapor phase
  infiltration of aluminum oxide into benzocyclobutene-based polymer
  dielectrics to increase adhesion strength to thin film metal interconnects.
  {\it Journal of Vacuum Science {\&} Technology A: Vacuum, Surfaces, and
  Films} 2020\string; 38(3)\string: 033210.
\newblock \href {\doibase 10.1116/1.5141475} {doi: 10.1116/1.5141475}

\bibitem{Ingram.2019}
Ingram WF, Jur JS. Properties and Applications of Vapor Infiltration into
  Polymeric Substrates. {\it JOM} 2019\string; 71(1)\string: 238--245.
\newblock \href {\doibase 10.1007/s11837-018-3157-9} {doi:
  10.1007/s11837-018-3157-9}

\bibitem{Chen.2021}
Chen X, Wu L, Yang H, Qin Y, Ma X, Li N. Tailoring the Microporosity of
  Polymers of Intrinsic Microporosity for Advanced Gas Separation by Atomic
  Layer Deposition. {\it Angewandte Chemie (International ed. in English)}
  2021\string; 60(33)\string: 17875--17880.
\newblock \href {\doibase 10.1002/anie.202016901} {doi: 10.1002/anie.202016901}

\bibitem{Liu.2022}
Liu Y, McGuinness EK, Jean BC, et al. Vapor-Phase Infiltration of Polymer of
  Intrinsic Microporosity 1 (PIM-1) with Trimethylaluminum (TMA) and Water: A
  Combined Computational and Experimental Study. {\it The Journal of Physical
  Chemistry B} 2022\string; 126(31)\string: 5920--5930.
\newblock \href {\doibase 10.1021/acs.jpcb.2c01928} {doi:
  10.1021/acs.jpcb.2c01928}

\bibitem{McGuinness.2019}
McGuinness EK, Zhang F, Ma Y, Lively RP, Losego MD. Vapor Phase Infiltration of
  Metal Oxides into Nanoporous Polymers for Organic Solvent Separation
  Membranes. {\it Chemistry of Materials} 2019\string; 31(15)\string:
  5509--5518.
\newblock \href {\doibase 10.1021/acs.chemmater.9b01141} {doi:
  10.1021/acs.chemmater.9b01141}

\bibitem{Dandley.2014}
Dandley EC, Needham CD, Williams PS, Brozena AH, Oldham CJ, Parsons GN.
  Temperature-dependent reaction between trimethylaluminum and poly(methyl
  methacrylate) during sequential vapor infiltration: experimental and ab
  initio analysis. {\it J. Mater. Chem. C} 2014\string; 2(44)\string:
  9416--9424.
\newblock \href {\doibase 10.1039/C4TC01293C} {doi: 10.1039/C4TC01293C}

\bibitem{Biswas.2014}
Biswas M, Libera JA, Darling SB, Elam JW. New Insight into the Mechanism of
  Sequential Infiltration Synthesis from Infrared Spectroscopy. {\it Chemistry
  of Materials} 2014\string; 26(21)\string: 6135--6141.
\newblock \href {\doibase 10.1021/cm502427q} {doi: 10.1021/cm502427q}

\bibitem{Dementyev.2023}
Dementyev P, Khayya N, Zanders D, Ennen I, Devi A, Altman EI. Size and Shape
  Exclusion in 2D Silicon Dioxide Membranes. {\it Small (Weinheim an der
  Bergstrasse, Germany)} 2023\string; 19(9)\string: e2205602.
\newblock \href {\doibase 10.1002/smll.202205602} {doi: 10.1002/smll.202205602}

\bibitem{Naberezhnyi.2022}
Naberezhnyi D, Mai L, Doudin N, et al. Molecular Permeation in Freestanding
  Bilayer Silica. {\it Nano letters} 2022\string; 22(3)\string: 1287--1293.
\newblock \href {\doibase 10.1021/acs.nanolett.1c04535} {doi:
  10.1021/acs.nanolett.1c04535}

\bibitem{Deilmann.2008b}
Deilmann M, Theiß S, Awakowicz P. Pulsed microwave plasma polymerization of
  silicon oxide ﬁlms: Application of efficient permeation barriers on
  polyethylene terephthalate. {\it Surface and Coatings Technology}
  2008\string; 202(10)\string: 1911-1917.
\newblock \href {\doibase https://doi.org/10.1016/j.surfcoat.2007.08.034} {doi:
  https://doi.org/10.1016/j.surfcoat.2007.08.034}

\bibitem{Steves.2013}
Steves S, Ozkaya B, Liu CN, et al. Silicon oxide barrier films deposited on PET
  foils in pulsed plasmas: influence of substrate bias on deposition process
  and film properties. {\it Journal of Physics D: Applied Physics} 2013\string;
  46(8)\string: 084013.
\newblock \href {\doibase 10.1088/0022-3727/46/8/084013} {doi:
  10.1088/0022-3727/46/8/084013}

\bibitem{Hegemann.2021}
Hegemann D, Bülbül E, Hanselmann B, Schütz U, Amberg M, Gaiser S. Plasma
  polymerization of hexamethyldisiloxane: Revisited. {\it Plasma Processes and
  Polymers} 2021\string; 18(2)\string: 2000176.
\newblock \href {\doibase https://doi.org/10.1002/ppap.202000176} {doi:
  https://doi.org/10.1002/ppap.202000176}

\bibitem{Starostin.2015}
Starostin SA, Creatore M, Bouwstra JB, Sanden v.~dMCM, Vries dHW. Towards
  Roll-to-Roll Deposition of High Quality Moisture Barrier Films on Polymers by
  Atmospheric Pressure Plasma Assisted Process. {\it Plasma Processes and
  Polymers} 2015\string; 12(6)\string: 545-554.
\newblock \href {\doibase https://doi.org/10.1002/ppap.201400194} {doi:
  https://doi.org/10.1002/ppap.201400194}

\bibitem{Elam.2017}
Elam FM, Starostin SA, Meshkova AS, et al. Atmospheric pressure roll-to-roll
  plasma enhanced CVD of high quality silica-like bilayer encapsulation films.
  {\it Plasma Processes and Polymers} 2017\string; 14(7)\string: 1600143.
\newblock \href {\doibase https://doi.org/10.1002/ppap.201600143} {doi:
  https://doi.org/10.1002/ppap.201600143}

\bibitem{Ozkaya.2015}
Ozkaya B, Mitschker F, Ozcan O, Awakowicz P, Grundmeier G. Inhibition of
  Interfacial Oxidative Degradation During
  SiO{\textless}sub{\textgreater}x{\textless}/sub{\textgreater}Plasma Polymer
  Barrier Film Deposition on Model Organic Substrates. {\it Plasma Processes
  and Polymers} 2015\string; 12(4)\string: 392--397.
\newblock \href {\doibase 10.1002/ppap.201400105} {doi: 10.1002/ppap.201400105}

\bibitem{Mitschker.2015}
Mitschker F, Dietrich J, Ozkaya B, et al. Spectroscopic and Microscopic
  Investigations of Degradation Processes in Polymer Surface-Near Regions
  During the Deposition of SiOx Films. {\it Plasma Processes and Polymers}
  2015\string; 12(9)\string: 1002-1009.
\newblock \href {\doibase https://doi.org/10.1002/ppap.201500085} {doi:
  https://doi.org/10.1002/ppap.201500085}

\bibitem{Mitschker.2017}
Mitschker F, Steves S, Gebhard M, et al. Influence of PE-CVD and PE-ALD on
  defect formation in permeation barrier films on PET and correlation to atomic
  oxygen fluence. {\it Journal of Physics D: Applied Physics} 2017\string;
  50(23)\string: 235201.
\newblock \href {\doibase 10.1088/1361-6463/aa6e28} {doi:
  10.1088/1361-6463/aa6e28}

\bibitem{Mitschker.2018b}
Mitschker F, Wißing J, Hoppe C, Arcos d.~lT, Grundmeier G, Awakowicz P.
  Influence of average ion energy and atomic oxygen flux per Si atom on the
  formation of silicon oxide permeation barrier coatings on PET. {\it Journal
  of Physics D: Applied Physics} 2018\string; 51(14)\string: 145201.
\newblock \href {\doibase 10.1088/1361-6463/aab1dd} {doi:
  10.1088/1361-6463/aab1dd}

\bibitem{Petasch.1997}
Petasch W, Räuchle E, Muegge H, Muegge K. Duo-Plasmaline — a linearly
  extended homogeneous low pressure plasma source. {\it Surface and Coatings
  Technology} 1997\string; 93(1)\string: 112-118.
\newblock \href {\doibase https://doi.org/10.1016/S0257-8972(97)00015-7} {doi:
  https://doi.org/10.1016/S0257-8972(97)00015-7}

\bibitem{Raeuchle.1998}
{R\"auchle, E.} . Duo-plasmaline, a surface wave sustained linearly extended
  discharge. {\it J. Phys. IV France} 1998\string; 08\string: Pr7-99-Pr7-108.
\newblock \href {\doibase 10.1051/jp4:1998708} {doi: 10.1051/jp4:1998708}

\bibitem{Hegemann.2007}
Hegemann D, Hossain MM, Körner E, Balazs DJ. Macroscopic Description of Plasma
  Polymerization. {\it Plasma Processes and Polymers} 2007\string; 4(3)\string:
  229-238.
\newblock \href {\doibase https://doi.org/10.1002/ppap.200600169} {doi:
  https://doi.org/10.1002/ppap.200600169}

\bibitem{kemaneci_2019}
Kemaneci E, Mitschker F, Benedikt J, Eremin D, Awakowicz P, Brinkmann {\relax
  R.P}. A numerical analysis of a microwave induced coaxial surface wave
  discharge fed with a mixture of oxygen and hexamethyldisiloxane for the
  purpose of deposition. {\it Plasma Sources Sci. Technol.} 2019\string;
  28\string: 115003.
\newblock \href {\doibase 10.1088/1361-6595/ab3f8a} {doi:
  10.1088/1361-6595/ab3f8a}

\bibitem{Eremin2022a}
Eremin D, Kemaneci E, Matsukuma M, Mussenbrock T, Brinkmann {\relax R.P}.
  Modeling of very high frequency large-electrode capacitively coupled plasmas
  with a fully electromagnetic particle-in-cell code. {\it Plasma Sources Sci.
  Technol.} 2023\string; 32\string: 044007.
\newblock \href {\doibase 10.1088/1361-6595/accecb} {doi:
  10.1088/1361-6595/accecb}

\bibitem{kemaneci_2017}
Kemaneci E, Mitschker F, Rudolph M, et al. A global model of cylindrical and
  coaxial surface-wave discharges. {\it J. Phys. D: Appl. Phys.} 2017\string;
  50\string: 245203.
\newblock \href {\doibase 10.1088/1361-6463/aa7093} {doi:
  10.1088/1361-6463/aa7093}

\bibitem{Eremin2022}
Eremin D. An energy- and charge-conserving electrostatic implicit
  particle-in-cell algorithm for simulations of collisional bounded plasmas.
  {\it J. Comput. Phys.} 2022\string; 452\string: 110934.
\newblock \href {\doibase 10.1016/j.jcp.2021.110934} {doi:
  10.1016/j.jcp.2021.110934}

\bibitem{mertmann_2011}
Mertmann P, Eremin D, Mussenbrock T, Brinkmann R, Awakowicz P. Fine-sorting
  one-dimensional particle-in-cell algorithm with {M}onte-{C}arlo collisions on
  a graphics processing unit. {\it Comput. Phys. Commun.} 2011\string;
  182(10)\string: 2161.
\newblock \href {\doibase 10.1016/j.cpc.2011.05.012} {doi:
  10.1016/j.cpc.2011.05.012}

\bibitem{turner_2013}
Turner {\relax M.M}, Derzsi A, Donkó Z, et al. Simulation benchmarks for
  low-pressure plasmas: Capacitive discharges. {\it Phys. Plasmas} 2013\string;
  20\string: 013507.
\newblock \href {\doibase 10.1063/1.4775084} {doi: 10.1063/1.4775084}

\bibitem{charoy_2019}
Charoy T, Boeuf J, Bourdon A, et al. 2D axial-azimuthal Particle-In-Cell
  benchmark for low-temperature partially magnetized plasmas. {\it Plasma
  Sources Sci. Technol.} 2019\string; 28\string: 105010.
\newblock \href {\doibase 10.1088/1361-6595/ab46c5} {doi:
  10.1088/1361-6595/ab46c5}

\bibitem{villafana_2021}
Villafana W, Petronio F, Denig A, et al. 2D radial-azimuthal Particle-In-Cell
  benchmark for $\mathbf{E} \times \mathbf{B}$ discharges. {\it Plasma Sources
  Sci. Technol.} 2021\string; 30\string: 075002.
\newblock \href {\doibase 10.1088/1361-6595/ac0a4a} {doi:
  10.1088/1361-6595/ac0a4a}

\bibitem{Berger2022}
Berger B, Eremin D, Oberberg M, et al. Electron dynamics in planar radio
  frequency magnetron plasmas: {III}. {C}omparison of experimental
  investigations of power absorption dynamics to simulation results. {\it
  Plasma Sources Sci. Technol.} 2023\string; 32\string: 045009.
\newblock \href {\doibase 10.1088/1361-6595/acc480} {doi:
  10.1088/1361-6595/acc480}

\bibitem{naberezhnyi2022}
Naberezhnyi D, Mai L, Doudin N, et al. Molecular Permeation in Freestanding
  Bilayer Silica. {\it Nano Letters} 2022\string; 22(3)\string: 1287-1293.
\newblock PMID: 35044780\href {\doibase 10.1021/acs.nanolett.1c04535} {doi:
  10.1021/acs.nanolett.1c04535}

\bibitem{dementyev2022}
Dementyev P, Khayya N, Zanders D, Ennen I, Devi A, Altman EI. Size and Shape
  Exclusion in 2D Silicon Dioxide Membranes. {\it Small} 2022\string: 2205602.
\newblock \href {\doibase doi.org/10.1002/smll.202205602} {doi:
  doi.org/10.1002/smll.202205602}

\bibitem{delosArcos.2022}
{de los Arcos} T, {C. Weinberger} , {F. Zysk} , et al. {Challenges in the
  interpretation of gas core levels for the determination of gas-solid
  interactions within dielectric porous films by ambient pressure XPS}. {\it
  {Applied Surface Science}} 2022\string; 604\string: 154525.
\newblock \href {\doibase 10.1016/j. apsusc.2022.154525} {doi: 10.1016/j.
  apsusc.2022.154525}

\bibitem{delosArcos.2023}
{de los Arcos} T, M{\"u}ller H, Weinberger C, Grundmeier G. {UV-enhanced
  environmental charge compensation in near ambient pressure XPS}. {\it
  {Journal of Electron Spectroscopy and Related Phenomena}} 2023\string;
  264\string: 147317.
\newblock \href {\doibase 10.1016/j.elspec.2023.147317} {doi:
  10.1016/j.elspec.2023.147317}

\bibitem{Richters2014}
Richters D, Kühne TD. Self-consistent field theory based molecular dynamics
  with linear system-size scaling. {\it J. Chem. Phys.} 2014\string;
  140(13)\string: 134109.
\newblock \href {\doibase 10.1063/1.4869865} {doi: 10.1063/1.4869865}

\bibitem{Schade2022}
Schade R, Kenter T, Elgabarty H, et al. Towards electronic structure-based
  ab-initio molecular dynamics simulations with hundreds of millions of atoms.
  {\it Parallel Computing} 2022\string; 111\string: 102920.
\newblock \href {\doibase 10.1016/j.parco.2022.102920} {doi:
  10.1016/j.parco.2022.102920}

\end{thebibliography}

\end{document}